\documentclass[10pt,journal]{IEEEtran}

\usepackage{tabularx}
\usepackage{textcomp}
\usepackage{booktabs} 
\usepackage{enumitem}
\usepackage{amsmath}
\usepackage{url}
\usepackage{graphicx}
\usepackage{subfigure}
\usepackage[table,xcdraw]{xcolor}
\usepackage{adjustbox}
\usepackage{array}
\usepackage{multirow}
\usepackage{makecell}
\usepackage{balance}
\usepackage{amsmath}
\usepackage{colortbl}
\usepackage{tcolorbox}
\usepackage{soul}
\usepackage[normalem]{ulem}
 \newcommand{\ciao}[1]{{\setlength\fboxrule{0pt}\fbox{\tcbox[colframe=black,colback=white,shrink tight,boxrule=0.2pt,extrude by=0.5mm]{\small #1}}}}

\newcommand{\ie}{{\em i.e., }}
\newcommand{\eg}{{\em e.g., }}
\newcommand{\myverb}{\fontsize{10}{48}\usefont{OT1}{lmtt}{b}{n}\noindent }

\ifCLASSINFOpdf
\else
\fi

\begin{document}

\title{Optimal Witnessing of Healthcare IoT Data \\Using Blockchain Logging Contract}

\author{Mohammad~Hossein~Chinaei,
	Hassan~Habibi~Gharakheili,
	and~Vijay~Sivaraman
	\IEEEcompsocitemizethanks{\IEEEcompsocthanksitem M. H. Chinaei, H. Habibi Gharakheili, and V. Sivaraman are with the School 
		of Electrical Engineering and Telecommunications, University of New South Wales, Sydney, NSW 2052, Australia (e-mails: m.chinaei@unsw.edu.au, h.habibi@unsw.edu.au, vijay@unsw.edu.au).
	}
}


\IEEEtitleabstractindextext{%
\begin{abstract}
Verification of data generated by wearable sensors is increasingly becoming of concern to health service providers and insurance companies. 
There is a need for a verification framework that various authorities can request a verification service for the local network data of a target IoT device.
In this paper, we leverage blockchain as a distributed platform to realize an on-demand verification scheme. This allows authorities to  automatically transact with connected devices for witnessing services. A public request is made for witness statements on the data of a target IoT that is transmitted on its local network, and subsequently, devices (in close vicinity of the target IoT) offer witnessing service. 

Our contributions are threefold: (1) We develop a system architecture based on blockchain and smart contract that enables authorities to dynamically avail a verification service for data of a subject device from a distributed set of witnesses which are willing to provide (in a privacy-preserving manner) their local wireless measurement in exchange of monetary return; (2) We then develop a method to optimally select witnesses in such a way that the verification error is minimized subject to monetary cost constraints;
(3) Lastly, we evaluate the efficacy of our scheme using real Wi-Fi session traces collected from a five-storeyed building with more than thirty access points, representative of a hospital. According to the current pricing schedule of  the Ethereum public blockchain, our scheme enables healthcare authorities to verify data transmitted from a typical wearable device with the verification error of the order 
0.01\% at cost of less than two dollars for one-hour witnessing service.
\end{abstract}

}

\maketitle

\IEEEdisplaynontitleabstractindextext

%
\IEEEpeerreviewmaketitle

\section{Introduction}
\indent
\IEEEPARstart{M}{odern } healthcare systems are increasingly coming online.  Wearable devices have profoundly improved patients experience of interacting with remote healthcare providers. They facilitate remote healthcare monitoring given their capability in automatic measurement of medical signs and periodic transmission of data over the Internet. Received medical data will be stored on a cloud server where different authorities can access data and take appropriate actions. Healthcare systems benefit from wearable devices to provide a higher quality of care to citizens, improved decision making, accurate real-time diagnosis, and timely treatment at considerably lower prices \cite{10waysiotmedi,iotmarket}. Many insurance companies have also developed policies based on the customers medical information received from wearable IoTs \cite{insurancereward}. 

Automated and online health services driven by new wearable technologies, while revolutionizing the traditional health systems, come at a price of frequent and sophisticated cyber threats  \cite{Kaspersky2017,IVT2017, pwc2018}. 
Wearable device users, themselves, sometimes become potential attackers when they try to forge data transmitted from their health sensor to claim pecuniary benefits \cite{siddiqi2016timestamp}.
An adversary, even located at far physical distance from a target wearable device, can maliciously manipulate data packets transmitted from the sensor to falsify its sensitive information. Attackers may attempt to manipulate the data transmitted from a health sensor such as an insulin pump, causing the patient to receive a lethal dose of medicine \cite{hackernews2017}. More worryingly, most of these attacks would remain undetected as neither the victim device nor the healthcare server is aware of an adversary in the middle of their communications. The largely scattered distribution of IoT-dependent patients coupled with poor security measures embedded in devices make wearables even more appealing victims to attackers who target health data authenticity and integrity. 
This poses massive risks to the entire healthcare data-centric system where practitioners medical decisions and many other analytical processes would rely solely on the accuracy and validity of data  measured by wearable sensors and transmitted via the Internet.

Any discrepancy in data,  between what is transmitted by the sensor and what is received by the healthcare system, warrants further investigation to check whether data is forged or tampered with.
Therefore, obtaining the local version of data (\ie transmitted by wireless sensors) is crucial for verification.
Neighboring nodes (those sharing a wireless broadcast domain with the target IoT sensor) can potentially provide a local version of its transmitted data -- we call these neighboring nodes as ``witnesses''. 
The witnesses are wireless devices that overhear data packets transmitted by the IoT sensor, and hence can record the local version of the data whenever needed (on-demand witnessing). The concept of crowdsourced secure logging for wearable devices was first proposed in \cite{siddiqi2019secure}, where wireless IoT sensors in an environment (within close proximity) statically overhear and record a fingerprint of other sensors' data, allowing a forensic expert to verify data transmitted from the sensors retrospectively. While proposing a novel idea, their scheme suffers from a number of issues including: (a) local network gateway presents a single point of failure, and hence difficult to scale; (b) the identity of witnesses is revealed, raising privacy concerns; and (c) witnesses lack incentives to statically participate in an opportunistic collection of witness records. To address these shortcomings we develop a dynamic scheme that allows authorities to publicly request for witnessing of data transmitted by a target sensor, and incentivized witnesses to collect and submit their local records in a privacy-preserving manner. 
In order to develop a practical scheme for data witnessing and verification, certain requirements are to be met:


\textbf{Secure and private logging:}
Witnesses record data packets from the target device and securely log them into their statements transmitted to and stored in a tamper-proof database for an admissible auditing. More importantly, statements should protect the privacy of witnesses (not revealing the identity and location of witnesses).


\textbf{Dynamic witnessing:}
Witnessing incurs computing costs, and hence should be performed judiciously and dynamically as opposed to statically. The Health Service Providers (HSP) queries over a distributed service platform for witness statements corresponding to a target device whenever needed. The query is seen by potential witnesses that have subscribed to this service platform. 
Those witnesses which are in close vicinity of the target sensor may choose to contribute to this process by giving statements at certain resolution -- higher resolutions incur heavier computing costs.

\textbf{Monetization:}
A potential witness has little incentives to participate in a witnessing process, incurring power and computing costs, unless there is a monetary return. HSPs potentially have the ability to pay; however, they need a systematic method that allows them to make dynamic decisions in choosing the right witnesses which meet their requirements while aligning with their budget. The HSP would choose certain witnesses (from a set of available ones) that yield the best verification subject to a budget constraint, and this selection can vary dynamically depending on the availability of witnesses in the environment and their ability in generating statements of certain resolutions.


Most of the previous works in the area of witnessing \cite{accorsi2009safe} focused on the first requirement above (\ie secure logging) in a variety of scenarios, while the other two essentials have been overlooked or poorly addressed. In this paper we use blockchain technology as a platform that connects individual witnesses to health authorities, enabling both parties to communicate, interact, and transact dynamically. 
Blockchain is widely perceived as an immutable, distributed, and decentralized database which is automatically updated by transactions via  consensus (agreement) among participants (peers). 
Recently, blockchain in conjunction with automatic executable programs (aka ``smart contracts'' \cite{zheng2018blockchain}) has enabled a medium for trading assets between untrusted parties. The distributed untrusted parties can communicate, negotiate, and establish a contract to trade their assets/services over the blockchain without any third-party intervention, while gaining the same functionality with a decent amount of certainty compared to centralized networks \cite{christidis2016blockchains}. Providing an immutable, shared, and real-time ledger along with enabling trustless transactions among individual peers motivate us to develop our witnessing scheme on a blockchain platform.
This paper describes a novel witnessing solution that allows healthcare authorities to remotely and dynamically verify the data received from  health IoT sensors. 
Witnessing requests are publicly 
Thanks to the blockchain technology and smart contracts, the healthcare authorities could request from any wireless device in the vicinity of the IoT sensor to overhear and record the local version of the data transmitted wirelessly by the device. The witnesses could securely make their statements based on the packets they are overhearing from the IoT device to provide the witness statements. 
Smart contracts on top of the blockchain network would allow witnesses to trade their statements with the healthcare service provider (HSP) in a privacy preserved manner without relying on trusted third parties. As a result, the HSP using witness statements would be able to find out any discrepancies between the data delivered to the healthcare cloud and the local version. 

Our contributions are threefold: (1) We develop a system architecture based on blockchain and smart contract that enables authorities to dynamically avail a verification service for data of a subject device from a distributed set of witnesses which are willing to provide (in a privacy-preserving manner) their local wireless measurement in exchange of monetary return; (2) We then develop a method to optimally select witnesses in such a way that the verification error is minimized subject to monetary cost constraints;
(3) Lastly, we evaluate the efficacy of our scheme using real Wi-Fi session traces collected from a five-storeyed building with more than thirty access points, representative of a hospital. According to the current pricing schedule of  the Ethereum public blockchain, our scheme enables healthcare authorities to verify data transmitted from a typical wearable device with the verification error of the order 
0.01\% at cost of less than two dollars for one-hour witnessing service.

The rest of the paper is organized as follows:  \S\ref{sec:prior} describes prior work on witnessing, data logging, and blockchain use-cases in data provenance, and \S\ref{sec:arc} describes our solution approach that systematically and dynamically makes use of smart contracts for distributed witnessing. In \S\ref{sec:model} we describe our optimization framework to select witnesses, while in \S\ref{sec:eval} we evaluate the efficacy of our scheme via simulation. The paper is concluded in \S\ref{sec:conc}.

\section{Related Work}\label{sec:prior}

This paper lies at the intersection of three strands of research works including intrusion detection systems (IDS), secure logging, and application of blockchain/smart-contract technologies.

Firstly, we emphasize that our objective is not to develop intrusion detection (IDS) method for IoTs, but alarms of a network IDS can be used to trigger our process of data verification. Secondly, we look at secure logging methods to maintain audit trail and discuss witnessing protocols that are close to our work. Thirdly, we highlight the application of blockchain and smart contract technologies in data provenance, access control management, and how our scheme differs from existing works.

\subsection{IDS for IoTs}
Intrusion detection systems monitor network traffic and often look for signature of malicious in traditional networks defined as monitoring the network components to detect any security violation. The IDS enables the auditor to detect abnormal behaviors in a network or system, highlighting possible compromised devices or cyber-attacks \cite{denning1987intrusion}. IDS in the IoT context would be different from the traditional use cases as it deals with very low-power devices on a very large scale, which shares data among objects and users \cite{raza2013svelte}. The strategies for IDS in IoT could be categorized based on different attributes, namely, IDS placement strategies, detection methods, and security threats. Interested readers may want to refer to these comprehensive surveys of IDS in IoT context \cite{zarpelao2017survey, mitchell2014survey, gendreau2016survey}.

In \cite{IoTSP18}, we proposed a network-based IDS for IoTs using manufacturer usage descriptions (MUD) and software-defined network (SDN) paradigm to translate formal behavioral profiles of off-the-shelf sensors to static and dynamic flow rules that can be enforced at run-time by the network administrator. Any sensor which its traffic deviates from the formal behavior would be detected as a potential intruder. This scheme is categorized as centralized IDS, which uses a hybrid of anomaly/signature-based methods to detect volumetric attacks, including reflection/amplification, flooding, ARP spoofing, and port scanning.

Our proposed witnessing service for verifying healthcare data in this paper differentiates from the IDS for IoT literature in different aspects. First, our work is not to detect the intrusion but could benefit from the IDS as an alarm, which forces the health authorities to run data verification service. Our service is placed in the cloud of the health authorities, not in the network administrator. Second, our system is an on-demand service to detect any attacks on the integrity and authenticity of data, in other words. While the IDS literature discussed above to inspect the sensor's traffic patterns to detect abnormal behaviors, we want to detect any discrepancy between the version of data transmitted from the sensor and the version delivered to the health cloud.

\subsection{Secure logging}
Log data in traditional systems are designed to record any event that changes the state of the system and would provide authorities to reconstruct the chains of events. The audit records may be used as digital evidence. The auditing architecture traditionally consists of three primary roles: dedicated devices that capture audit records, collectors who store these records, and an auditor who retrieves the collected data and retroactively investigates the records to detect any suspicious activity. Various secure logging methods have been proposed in both capturing and storing phases to fulfill the requirements of logging audit records as admissible evidence \cite{accorsi2009safe}. In the following, we briefly discuss important works that address more security requirements.


In \cite{schneier1999secure}, authors consider a secure logging architecture in which the audit logs would be stored on an untrusted machine in the storage phase. They proposed a scheme based on hash chains and evolving shared cryptographic keys to limit the attacker's ability to read and alter the audit logs. The shared key in each epoch is derived from the hash of the previous data logs. Therefore, when an attacker could break one of the shared keys, he only would be able to change audit logs for that specific epoch. Any audit log manipulation would be detectable because the shared keys for the next epochs have been made using the original version of the audit log. The major drawback is that any verifier needs to posses the shared key of epoch to verify the authenticity of audit logs. This issue has been addressed by Logcrypt \cite{holt2006logcrypt}, where the author proposed using public key encryption, which enables the audit log to be signed by one entity and be verifiable to anyone else without revealing the secrets.   

Authors in \cite{ma2009new}, while pointing out the security attacks which the previous works are vulnerable to, propose a secure logging architecture based on Forward-Secure Sequential Aggregate (FssAgg) authentication techniques. Their proposed scheme takes a private key, a message to be signed, and the aggregated signature computed up to this point to compute a new aggregate signature. Their evaluation shows that this scheme provides better security while incurring less computational and communication overhead. All of the schemes above need either predefined powerful audit generators in the vicinity of the IoT sensor or very high computational power at the sensor itself to digitally sign its audit logs and transmit them to the authorities. None of them would be applicable in the IoT context due to the large scale of these energy-poor devices.

\subsubsection{Witnessing} 
To the best of our knowledge, witnessing as an approach to seek assistance from the neighboring nodes has mostly been applied in the location proof systems. In \cite{zhu2011applaus}, authors propose APPLAUS a privacy-preserving location proof system that enables different verifiers to verify a node's location based on the proofs collected from the witnesses in the environment. All of the parties, including the prover node and witnesses, should register with a certification authority (CA) to obtain secrets for signing the proofs. Any node registered in the system and equipped with the Bluetooth technology could ask all available witnesses in the environment to provide him a location proof and forward these proofs to a central proof server where an authorized verifier could query and retrieve location proofs from. To protect real identities of different parties(nodes and witnesses) from each other and from location proof server, The CA uses a pseudonym approach that provides each user with a set of private/public key pairs. Authors in \cite{wang2013stamp} propose an improved scheme named STAMP based on a unique pair of public/private key and commitment techniques instead of periodically changed pseudonyms. They also remove the untrusted central proof server to perform better collusion detection.

In \cite{hasan2016woral}, authors propose WORAL as a witness oriented provenance framework for secure location proofs of the mobile devices. In this scenario, each physical region has a designated location authority, and a set of mobile users, which could play the role of volunteer witnesses, or a user needs to prove her location. Witnesses provide notarization of a statement between the user and location authorities of an environment. The devices have local storage for storing these provenance items, which is fully controlled by its user, which could partially or entirely provide them to different applications that asked for location proofs. All of the aforementioned papers are in the context of secure location proofs. In our scenario, we need the witnesses to provide health authorities with a compact local version of the data transmitted from the IoT sensor. We can not rely on the location authorities as it needs predefined static structures on a very large scale to cover all of the sensors.  On the other hand, we can not incur the computational and communication power of generating periodic signatures or commitment techniques to the ultra-low-power sensors—also, all of these methods lack a monetization scheme to incentivize the witnesses.


The authors in \cite{siddiqi2019secure} propose an algorithm of witnessing for secure logging and forensics of medical data. They assume that while a sensor is communicating with its associated gateway, the other sensors could overhear the data packets. The witnesses (neighboring sensors communicating with the same gateway) in the vicinity of a sensor log any communication packets occurred between the gateway and the sensor in their bloom filters as an efficient mechanism to keep the fingerprint of the sensor's data packets. At the end of the epoch, witnesses would upload their bloom filters in a central forensic server, and the gateway would send the list of witnesses to the forensic server.

Their scheme, while proposing a novel application of witnessing, suffers from severe vulnerabilities regarding security and implementation. First, their scheme would disclose the location of the gateway, witnesses, and the sensor to a third party, compromising the privacy of all entities contributing to the scheme. Second, this protocol suffers from a single point of failure since the gateway in the environment could easily cheat in sending the list of witnesses to the central server. Third, their scheme does not scale. It needs different authorities such as servers of the witnesses, Telco server, and the healthcare server, to pre-coordinate with each other. Four, the witnesses are not incentivized to contribute to the scheme.

\subsection{Blockchain}

The emerging blockchain technology and smart contract on top of it provide a shared, distributed database governed by a consensus algorithm, not a single authority. The majority of the network would decide upon accepting new events and chain them to the rest of the history (previous events). The cryptographic structure of the data stored on the blockchain and the enormous amount of power needed to reach an agreement between different peers makes the Blockchain tamper-proof against adversaries. The decentralized control and immutability of the blockchain make it a viable approach to develop trusted systems \cite{tschorsch2016bitcoin}. 

Smart contracts are pre-defined contractual terms written in the form of digital scripts which are executed if certain conditions are met. While traditional blockchains like bitcoin support only cryptocurrency transactions, more advanced Blockchain like Ethereum support smart contracts providing individual users with the opportunity to self-enforce their protocols to transfer different assets \cite{zheng2018blockchain}.  

Blockchain as a tamper-proof database has many capabilities to support data provenance, which defines as meta-data that keeps track of data transformation and its ownership from the source it is originated to its current shape \cite{blockusecase}. In \cite{liang2017provchain}, authors devise a scheme based on smart contracts to track the changes made to a document while it is stored on the cloud. It would record different parties' actions to change the data and store the records on the blockchain using smart contracts. The proposed architecture in \cite{ramachandran2017using} using on-chain voting smart contract protects the document against unauthorized users or changes that are not verified by the majority of authorized users.

Another use-case of Blockchain in the literature is for access control management. The shared distributed nature of the Blockchain would facilitate access control management. One of the best papers in this field is \cite{ouaddah2016fairaccess}, in which authors use the Blockchain technology to provide an access control framework for IoT devices. Their framework uses the transactions on the shared ledger to grant, get, revoke, and delegate access in the IoT applications. They emphasize on the rights of individual users as owner of the data and use the shared ledger of the Blockchain as an immutable, tamper-proof database which enables the user of the sensor to get access to different parties.     

We use Blockchain in a different context to reach different aims. Our scheme differs from the previous papers in the area of Blockchain as it develops a dynamic witnessing algorithm that allows witnesses colocated with the health wearable device to contribute and provide the healthcare system with the local version of the data. The local versions of the data would be considered an asset to the healthcare provider to discover security attacks on data integrity and authenticity. We have utilized smart contracts for trading multiple assets (witness statements and cryptocurrencies) on the Blockchain without reliance on an authorized third party.

\section{System Architecture And Algorithm}\label{sec:arc}
In this section, we develop our system architecture for an on-demand auditing of wearable data using distributed witness statements that are enabled by smart contracts. We first outline the threat model, then develop our witnessing system architecture, and finally describe the flow of events in an operational scenario.

\begin{figure}[!t]
	\centering
	\includegraphics[width=0.48\textwidth]{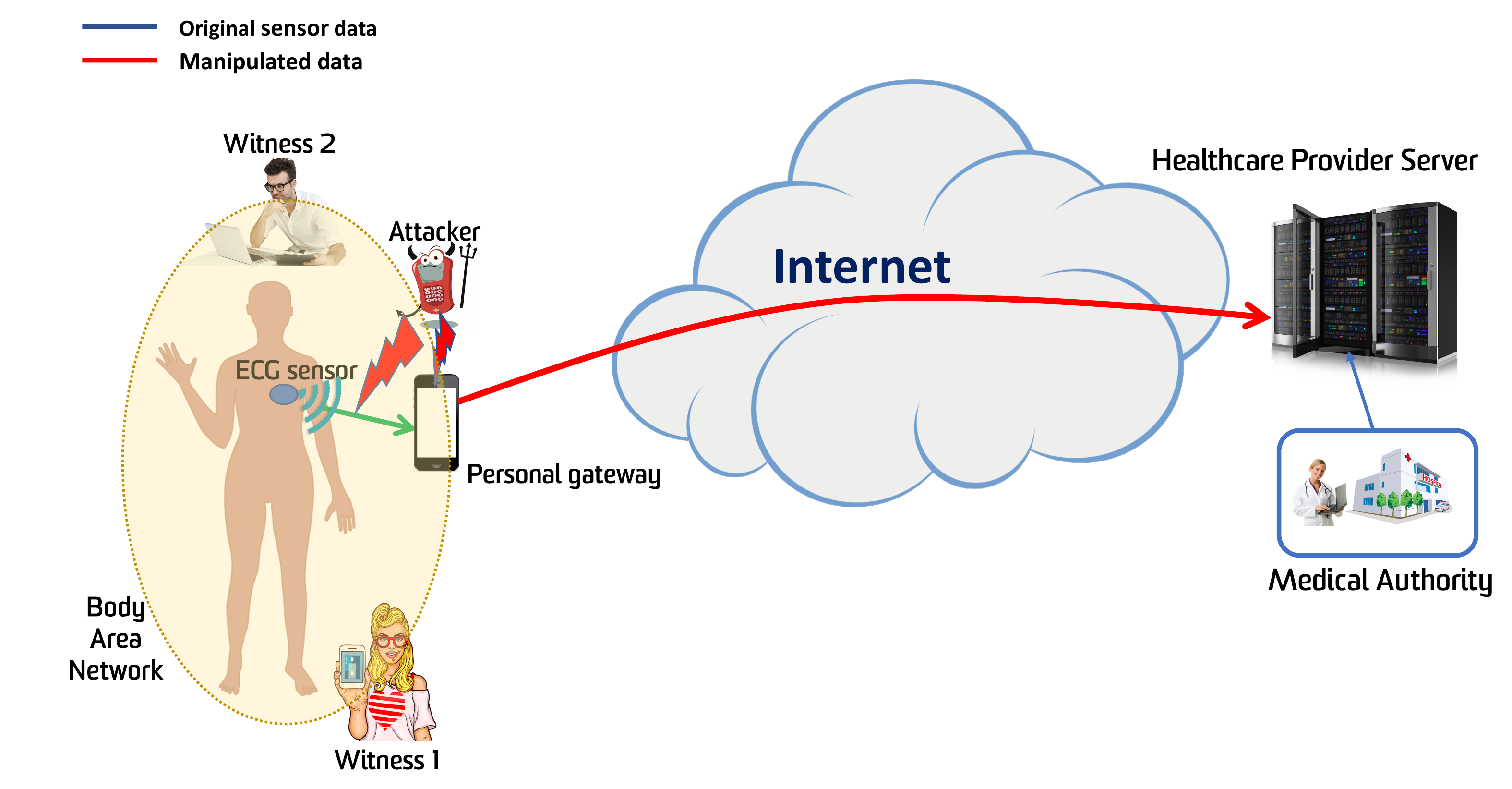}
	\vspace{-3mm}
	\caption{Threat model: forging IoT data in transit by man-in-the-middle.} 
	\label{fig:threat}
\end{figure}





\subsection{Threat Model}
Body area networks (BAN) consist of wearable sensors that measure vital signs of body and transmit measurements toward a personal gateway at which the data is collected and forwarded to the HSP through the Internet (Fig.~\ref{fig:threat}). In this paper, we focus on man-in-the-middle attackers that can manipulate IoT data in transit from wearable sensors to the personal gateway. The attacker could be either the owner of the sensor itself, falsifying the medical measurements in order to claim financial benefits, or other malicious entities who aim to harm to individual patients or the broader health-care system. 
In this environment: the attacker can compromise the gateway to forge the data measured and transmitted by the sensor; the attacker can backfill medical data \cite{siddiqi2016timestamp}; and  witnesses are independent and trusted entities which are assumed to be not compromised -- the HSP has the freedom to choose from a number of potential witnesses that are present in the environment. 

%

\subsection{System Architecture}
\begin{figure}[!t]
	\centering
	\includegraphics[width=0.48\textwidth]{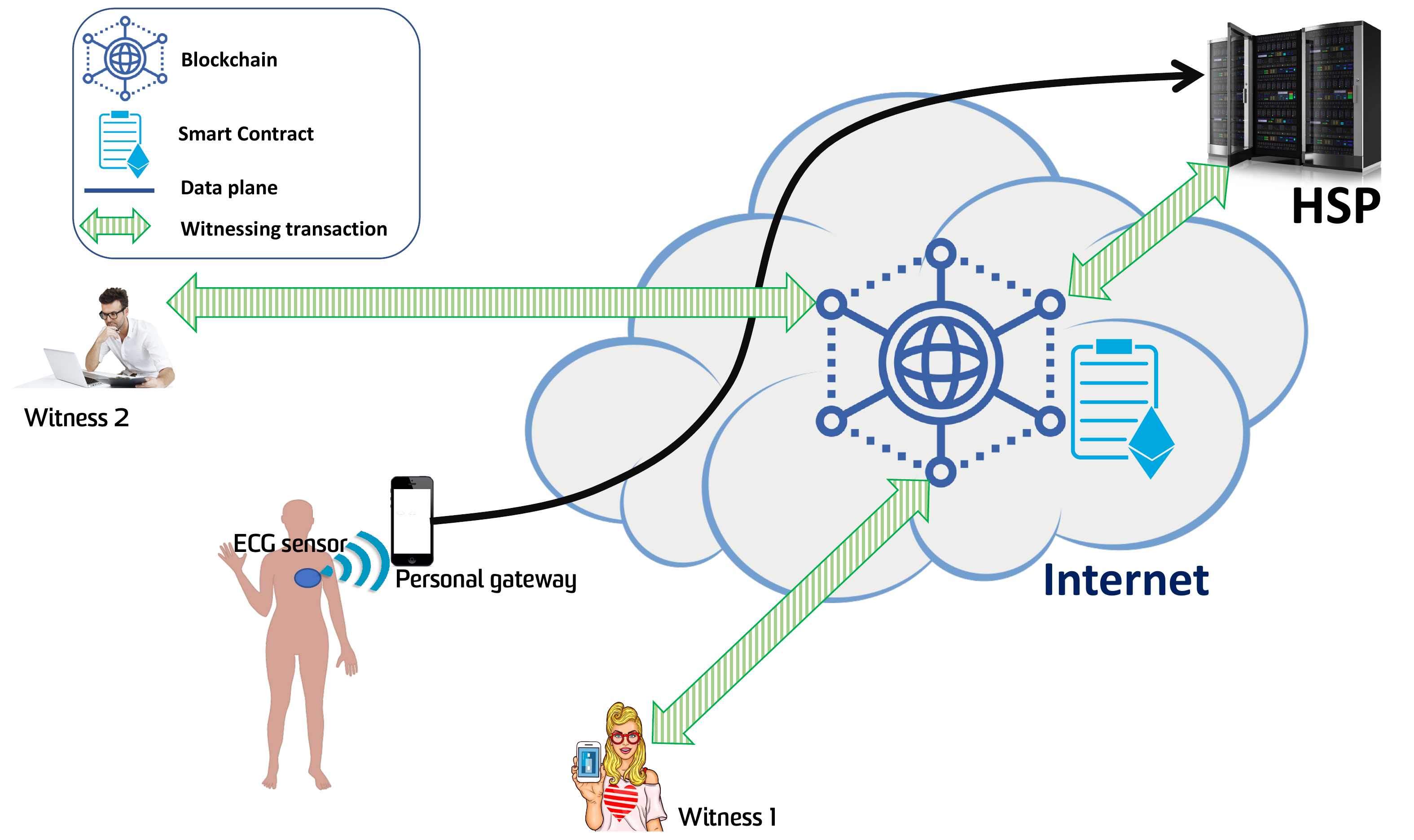}
	\vspace{-2mm}
	\caption{System architecture of on-demand witnessing over blockchain}
	\label{fig:arc}
\end{figure}

Fig.~\ref{fig:arc} shows the system architecture of our witnessing scheme. We now explain various entities in this architecture.

\textbf{Healthcare Service Provider (HSP)}
is an entity, shown on top right of Fig.~\ref{fig:arc}, that provides remote health services to patients based on the medical data received from their body-worn sensor. The HSP receives, stores, and manages the access to real-time medical data to provide timely support and treatment to the patients.

\textbf{Wearable IoT Sensor} 
is an on-body low-power sensor to measure physiological signs of the patient. The sensor transmits the health data to a personal gateway from where the data gets forwarded to a remote server of the HSP on the Internet.   


\textbf{Witnesses}, by their definition, are wireless nodes with Internet connectivity (\eg Witness1 and Witness2 in Fig.~\ref{fig:arc})  that locate in the physical vicinity of the wearable IoT sensor, and share the same wireless broadcast domain. Witnesses are able to overhear the data transmitted wirelessly by the wearable IoT. Since the wearable data is encrypted, witnesses are not able to extract any information from the overheard data packets.  

\textbf{Blockchain Network}
maintains a distributed public ledger, shared across peering nodes. Transactions between nodes get recorded in a chronological order, and are linked to previous ones by cryptographic mechanisms. All transactions within a specific time period are combined into a block which is chained (linked) to previous blocks, extending the ledger. A new block is added to the ledger only after it is approved by a consensus mechanism, \ie a majority of network nodes (aka \textit{miners}) verify its validity. 
Traditional blockchain networks like Bitcoin are typically used for financial transactions (exchange of digital money) \cite{tschorsch2016bitcoin}, while modern blockchains like Ethereum serve for a wider range of customized and programmable transactions (exchange of valuable tokens based on smart contracts) \cite{zheng2018blockchain}.
Nodes (the HSP and witnesses) can join a blockchain network by creating their account, consisting of  a pair of public and private keys -- the private key is used to sign transactions, and the public (their identity across the network) will be used for verification.


\textbf{Smart Contracts}
are special accounts that are created for specific application (\ie witnessing services) by a node (\ie the HSP in this paper), and become available to every nodes on the Ethereum network. Smart contracts come with a unique identifier, and typically offer a range of functions (\S\ref{sec:flow}) that can be called by a node which submits a transaction on the blockchain. The access to these functions can be controlled (\eg available to all nodes or restricted to certain nodes) by the node which develops and deploys the smart contract on the blockchain.  

\begin{figure}[!t]
	\centering
	\includegraphics[width=0.48\textwidth]{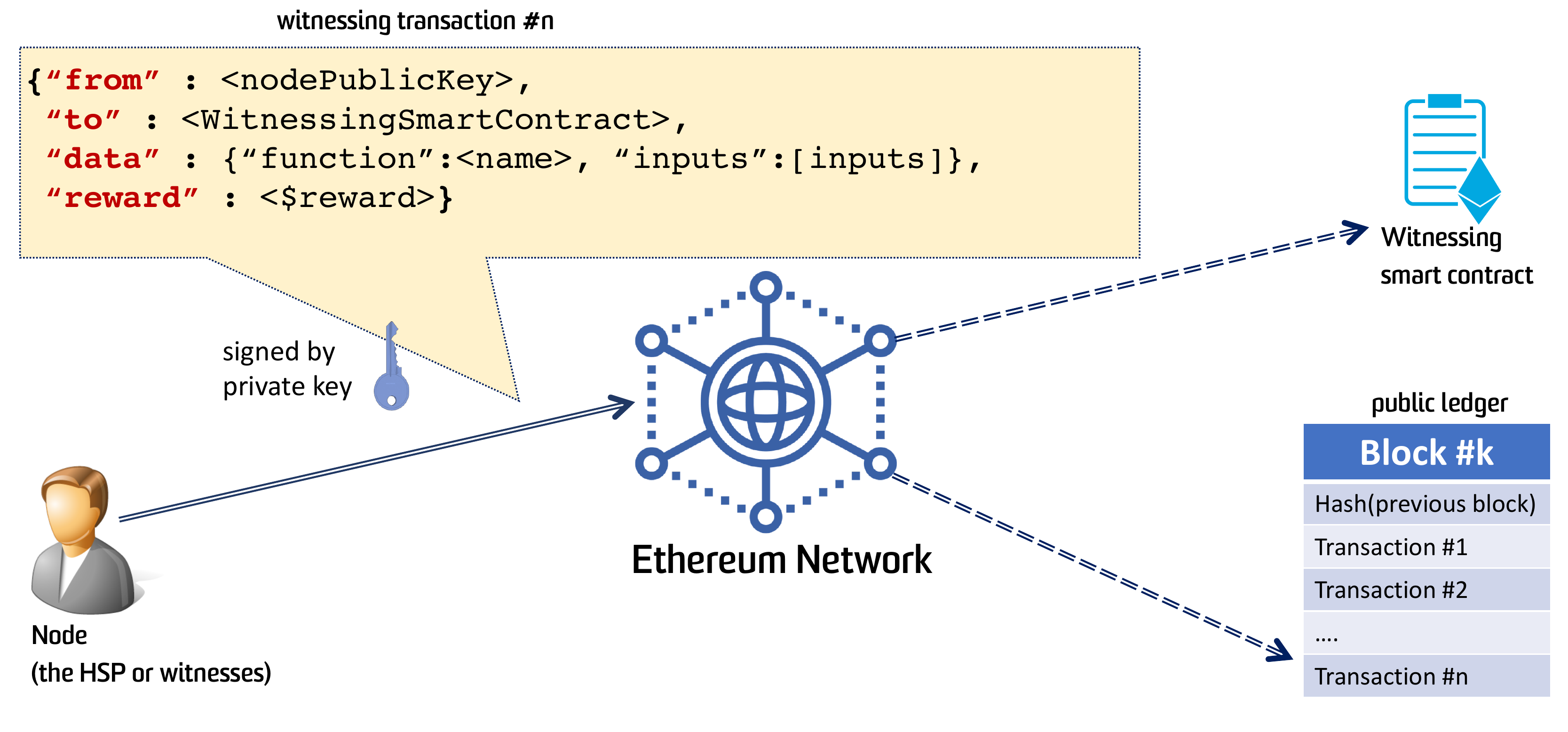}
	\vspace{-2mm}
	\caption{Witnessing transaction sent by nodes on Ethereum blockchain, updating the state of smart contract and appending the ledger.}
	\label{fig:transaction}
\end{figure}

Fig.~\ref{fig:transaction} shows how a witnessing transaction is made on the Ethereum network, updating the state of the  contract (depending on the specific function call) and ultimately getting appended to the ledger. Each transaction consists of four essential fields including: (a) {\myverb{"from"}} the identity of sender node, (b) {\myverb{"to"}} the identity of the witnessing smart contract, (c) {\myverb{"data"}} which contains of a {\myverb{"function"}} name along with a pair of {\myverb{"inputs"}} and {\myverb{"outputs"}} that vary by function (\S\ref{sec:flow}), and (d) {\myverb{"reward"}} offered by the caller to miners of the blockchain to approve the witnessing transaction and append it to the public ledger.    
We note that the HSP and potential witnesses on this network are particularly interested in the witnessing transactions (\ie addressed to the witnessing smart contract) of the ledger -- other types of transaction may occur on this blockchain network.




\begin{figure}[!t]
	\centering
	\includegraphics[width=0.465\textwidth]{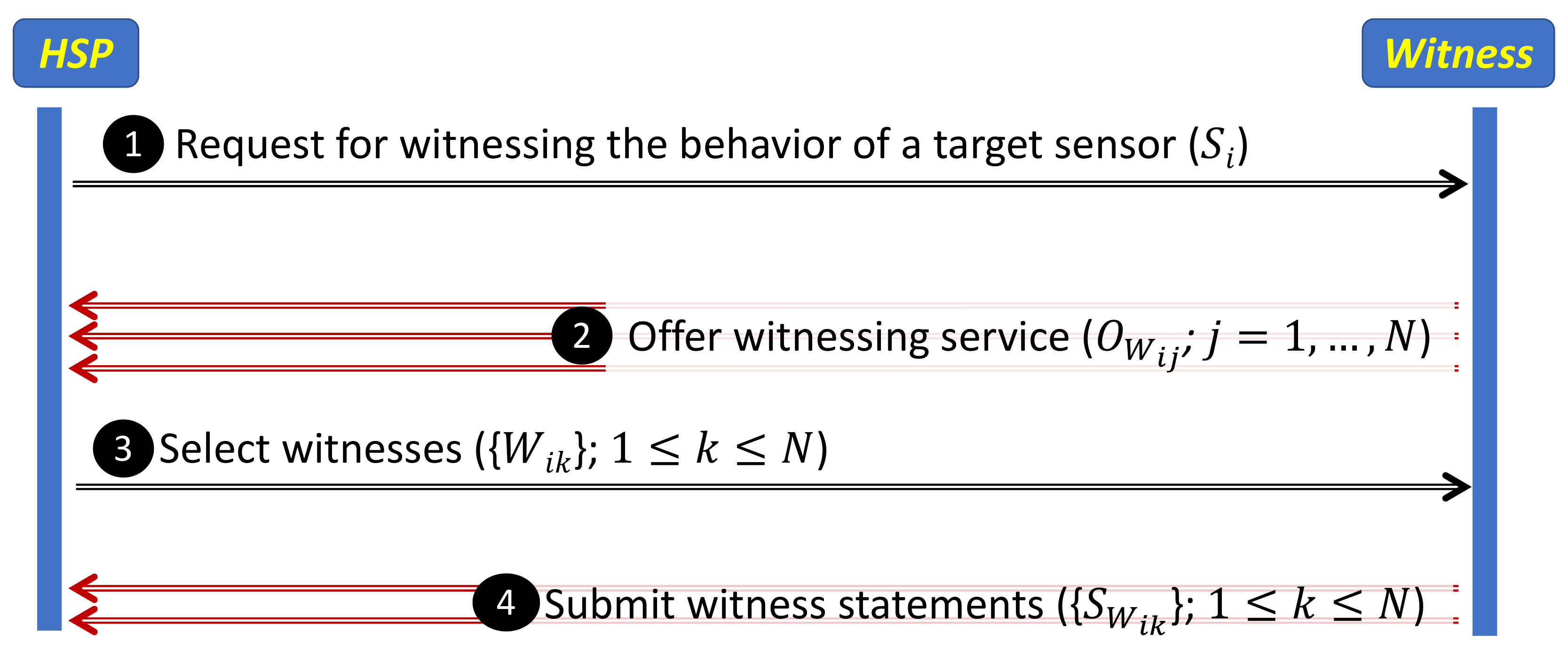}
	\vspace{-2mm}
	\caption{Sequence of witnessing interactions between the HSP and potential witnesses over blockchain.} 
	\vspace{-4mm}
	\label{fig:flowEvent}
\end{figure}


\subsection{Flow of Events}\label{sec:flow}
Wearable sensors transmit their measured data via their gateway to the HSP server on the Internet (shown by solid black line in Fig.~\ref{fig:arc}). 
Note that the data is sent over the wireless medium on the local network (\eg inside hospitals or homes) to the gateway, and hence all communications between gateway and wearable healthcare devices are overheard by other parties in the vicinity. 
Man-in-the-middle attackers (with different incentives) can intercept packets sent by the wearable devices and may attempt to tamper the data in transit. There are multiple adversary models \cite{andrea2015internet,yang2017survey} for manipulating data while being transmitted on the network: key management process (between the IoT device and HSP server) can be exploited by the man-in-the middle to access the secret key; an attacker may also collect previous packets of the IoT device and re-send them maliciously at later times to mount a replay attack on the HSP. 
For a given environment (say, a home), we have an IoT sensor ($S_i$) with a personal gateway (${GW}_i$), and multiple witnesses ($W_{ij}$).
Fig.~\ref{fig:flowEvent} illustrates the flow of events in a witnessing process that consists of a sequence of four steps explained as follows. 

\begin{figure}[t!]
	\centering
	\includegraphics[width=0.485\textwidth]{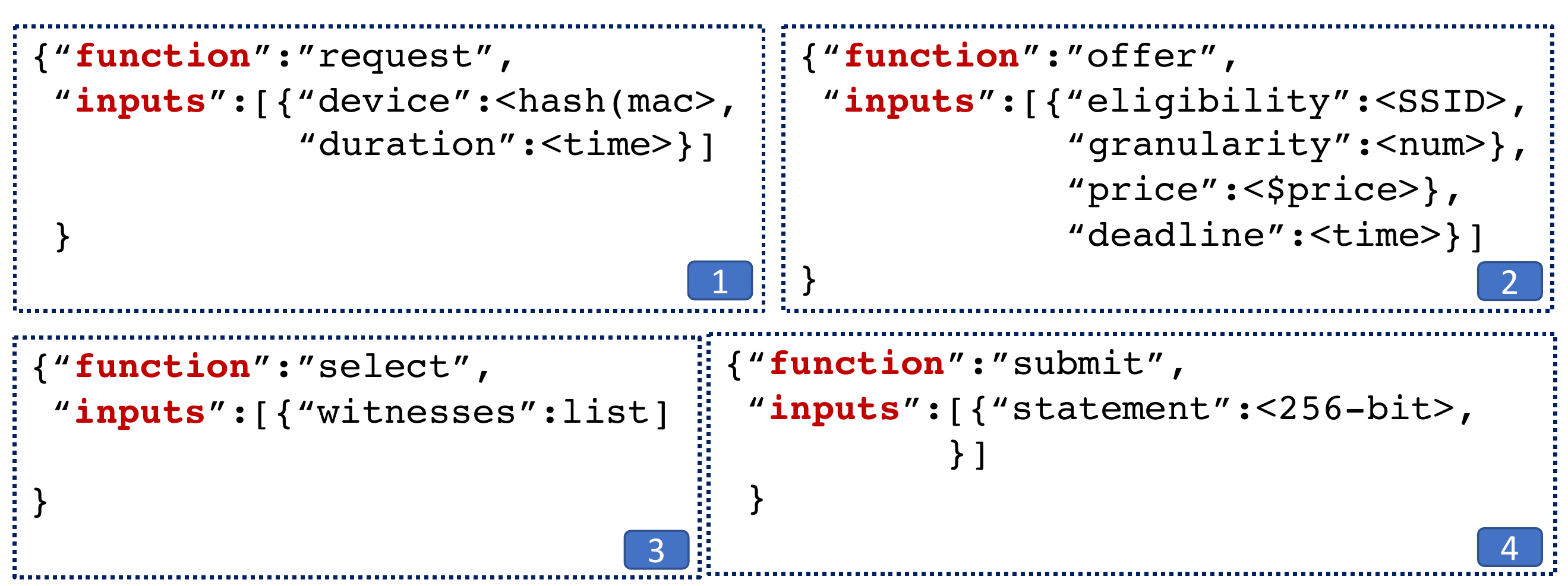}
	\vspace{-4mm}
	\caption{Components of data field in witnessing transactions vary by function type.} 
	\vspace{-4mm}
	\label{fig:datalog}
\end{figure}

\textbf{\ciao{Step1} Request for Witnessing:}
The HSP initiates the process by invoking a function (of the smart contract) called {\myverb{"request"}}, passing two parameters namely the identity of a target sensor (\eg {\myverb{"device":<hash(mac)>}}), and a desired duration of witnessing {\myverb{"duration":<time>}} for {\myverb{"inputs"}} component of the {\myverb{"data"}} field, as shown on the top left of Fig.~\ref{fig:datalog}. Note that for privacy reasons a hashed version of the sensor's MAC address ({\myverb{"hash(mac)"}}) is publicly announced on the network. Also, the {\myverb{"request"}} function can only be called by the HSP node, and potential witnesses will continuously look for this specific transaction submitted by the HSP.  
Upon arrival of a witnessing request, witness nodes will check their local network (wireless LAN) whether they are in the vicinity of the target sensor, or not. 
Note that the HSP would require witnesses to complete an eligibility challenge, proving that they are indeed in the vicinity of the target sensor location. The eligibility challenge has been widely studied in the literature \cite{hightower2000spoton,lazos2005rope,li2005robust,moore2004robust,capkun2005secure}, and is beyond the scope of this paper. That being said, wireless SSID association, public IP subnet, or \textit{nonce} packets emitted by the target sensor can be used as a loose proof of location. 


\textbf{\ciao{Step2} Offer Witnessing Service:}
Those nodes which find themselves local to the target sensor, may choose to offer the witnessing service depending upon their available resources like memory, battery, or compute power. Each potential witness will respond by invoking a function of the smart contract called {\myverb{"offer"}}, passing a response to the  {\myverb{"eligibility"}} challenge (\eg SSID), the  {\myverb{"granularity"}} of their statements (\eg number of packets to be included per statement), the dollar {\myverb{"cost"}} of their service, and (optional) a {\myverb{"deadline"}} by which they aim to make their statement available on the blockchain. Statements can only get added to the ledger if they are approved by a majority of the blockchain miners. We note that miners prioritize those transactions (block of transactions) which offer higher rewards, and hence the approval of less-rewarding transactions can get delayed.


The granularity of witness statements  depends on the resources such as memory, battery, and compute power that are available on the potential witness device for generating, transmitting, and more importantly submitting statements to the blockchain -- the higher the granularity, the more resources needed, and thus resulting in higher price (cost) of witnessing. The deadline of statement depends on the load of the blockchain network and the dollar incentive (reward) that the sender (witness) is willing to incur. We note that a congested network may delay the availability of statements on the ledger, unless a higher reward is paid. The witnessing price depends on the deadline metric plus an additional remuneration that individual witnesses may want to receive from the HSP for the service to be provided. This means that availing a high granular statement in a short time from a demanding witness, can be quite expensive for the HSP.

\textbf{\ciao{Step3} Witness Selection:}
Once the witnessing offers are submitted by the eligible witnesses, the HSP needs to select those which give the best quality statements subject to a limited (dollar) budget available.  
Note that the HSP aims to achieve a verification probability close to 1 for the data transmitted by its target sensor -- this objective is met when a large number of witnesses submit fine-grained statements. 
Considering the objective and constraint, the HSP selects a set of (zero, one or more) witnesses and announces them on the blockchain, committing to a dollar cost in exchange of the witnessing services to be provided by those selected witness nodes. This announcement will be made on the blockchain network by invoking {\myverb{"select"}} function of the witnessing smart contract. Again, the access to this function is restricted to the HSP as the authority in charge of witness selection.

\textbf{\ciao{Step4} Submit Witness Statement:}
Once a witness node is notified of its selection via the {\myverb{"select"}} transaction submitted by the HSP, it starts overhearing the packets transmitted by the target sensor and generates witness statements. Each of these statements is sent onto the Ethereum network by calling {\myverb{"submit"}} function of the smart contract. One this specific typs of transaction is approved by the network miners, the smart contract will automatically pay off \textit{Ethers} equal to the requested amount of {\myverb{"price"}} (in{\ciao{step3})  to the witnesses from the HSP's account.

\section{Witness Statements and \\Optimal Witness Selection} \label{sec:model}
Once the HSP ensures that potential witnesses are in the vicinity of its target sensor, its objective becomes to select a group of them that yield the best verification accuracy, given a limited budget. 
In this section we first describe the structure of witness statements, and then develop an optimization problem of selecting witnesses.

\subsection{Witness Statements and Verification}\label{sec:verification}   
We consider a standard form of witness statements (with flexible granularity) that fulfill the following requirements: (a) not revealing any information of the sensor data, (b) allowing the HSP to check whether a specific data packet has been logged into the statement with some degree of certainty, and (c) being lightweight for resource-constrained witnesses to participate.
One of the possible candidates for witness statement is bloom filter which was employed in \cite{siddiqi2019secure} to provide opportunistic binding of the medical data to its context. 

A bloom filter is a probabilistic data structure that is typically used to add elements (data packets) to a set (filter) and test if an element is in a set. Instead of  the elements themselves, a hash of them is added to the set. When testing if an element is in the bloom filter, \textit{false positives} are possible -- either an element is definitely not in the set or that it is possible the element is in the set.
An empty Bloom filter is a bit array of $M$ bits, all set to $0$. There are also $k$ independent hash functions, each of which maps an element to one of the $M$ bit positions. To add an element, feed it to the hash functions to get $k$ bit positions, and set the bits at these positions to $1$. Note that with more elements embedded in the filter, the error rate (false positive) increases. For given filter size $M$ and the probability of false-positive $f$, the number of data elements $n$ in the filter is determined by:

\begin{equation}
	n = \frac{-M (ln2)^{2}}{ln(f)} \label{bloom1}
\end{equation}

Also, the optimum number of hash functions needed to generate a bloom filter of size $M$ bits with $n$ logged elements is given by:
\begin{equation}
	k = \frac{M}{n}ln2 \label{bloom2}
\end{equation}

For our application, each witness commits to a number of inserted packets in each statement $n$ (passed as {\myverb{granularity}} in \ciao{step2}) while the size of bloom filters for all witness statements is fixed (say, $M=256$). Witnesses, therefore, may generate and submit a number (denoted by $m$) of statements based on their target false-positive probability $f$ and the total number of packets (denoted by $N$)  they have heard from the IoT sensor. 

\begin{equation}
	m = \frac{N}{n}=\frac{-Nln(f)}{M (ln(2))^{2}} \label{num_state}
\end{equation}

On the other hand, the HSP verifies the presence of certain packets in the submitted bloom filters by applying the same hash functions used to generate the statement. 
To test if a packet is in the filter, the HSP feeds it to the hash functions to get $k$ bit positions. If any of the bits at these positions is $0$, the packet definitely is not in the filter. If all are $1$, then the packet may be in the filter with the probability of $1-f$. 
It is important to note that a negative response from the statement is certain since bloom filter cannot result in a false negative. Since witnesses are independent, the verification probability (denoted by $\tau$) using bloom filters submitted by $W$ witnesses can be derived from: 

\begin{equation}
	\tau = 1 - \prod_{i=1}^{W}{f_{i}}  
\end{equation}

\subsection{Optimal Witness Selection}
As discussed in \S\ref{sec:flow} (\ciao{step2}) each potential witness offers a price for the service requested by the HSP. This price (denoted by $\alpha$) is the sum of a reward for blockchain miners and the remuneration expected by the potential witness. Note that witness statements need to be approved by the blockchain miners, and thereby get appended as a valid transaction to the public ledger.        
Obviously, targeting a lower false-positive probability $f$ requires a larger number of bloom filters (statements) to be submitted by the witness that results in a higher price charged to the HSP. Therefore, the cost of receiving witness statements from a witness which commits to error probability $f$ at price $\alpha$ is given by:
\begin{equation}
	c =m\alpha=\frac{-N ln(f)\alpha}{M (ln(2))^{2}} \label{costformula}
\end{equation}


Given a list of offers from $W$ potential witnesses, each with ($f_i$, $c_i$), the HSP needs to select a combination of them that collectively give the highest verification probability (the lowest error) subject to its budget constraint (denoted by $C$). This can be formally defined as an optimization problem:

\begin{equation}
	\begin{aligned}
		\max \quad & 1-\prod_{i=1}^{W}{(f_{i})^{x_{i}}}\\
		\textrm{s.t.} \quad & \sum_{i=1}^{W} x_{i}c_{i}\leq C\\
	\end{aligned}
\end{equation}

The objective function is maximized over $x_{i}$ which indicates whether the $i_{th}$ witness to be selected by the HSP.

\begin{equation}
	x_{i} =
	\begin{cases}
		1 &witness\;\textbf{\textit{i}}\;is\;selected,
		\\
		0 &witness\;\textbf{\textit{i}}\;is\;not\;selected,
		
	\end{cases}
\end{equation}

\begin{figure*}[t!]
	\begin{center}
		\mbox{
			\subfigure[Verification error.]{
				{\includegraphics[width=0.47\textwidth,height=0.30\textwidth]{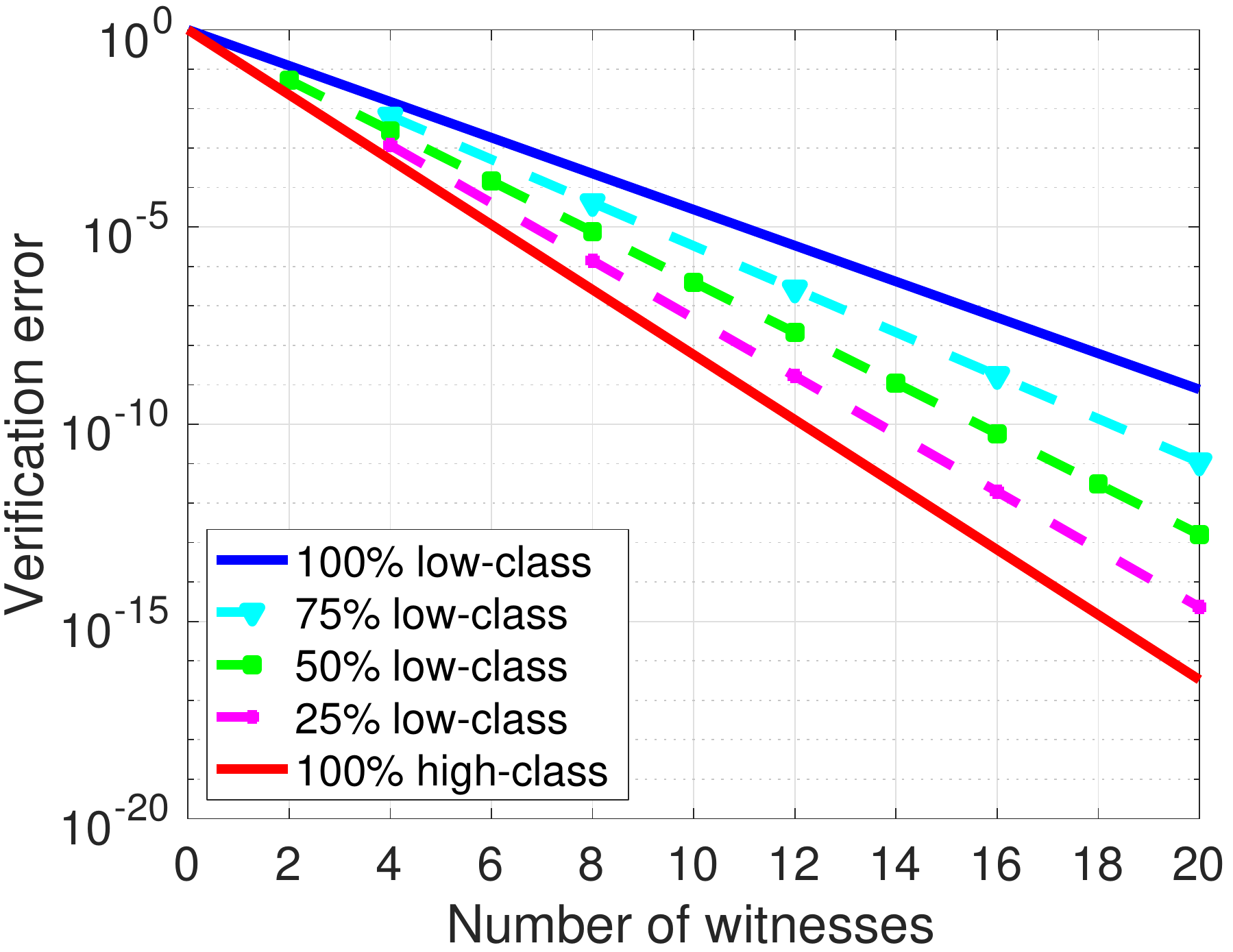}}\quad
				\label{fig:veritradeLowHigh}
			}
			\subfigure[Total cost.]{
				{\includegraphics[width=0.47\textwidth,height=0.30\textwidth]{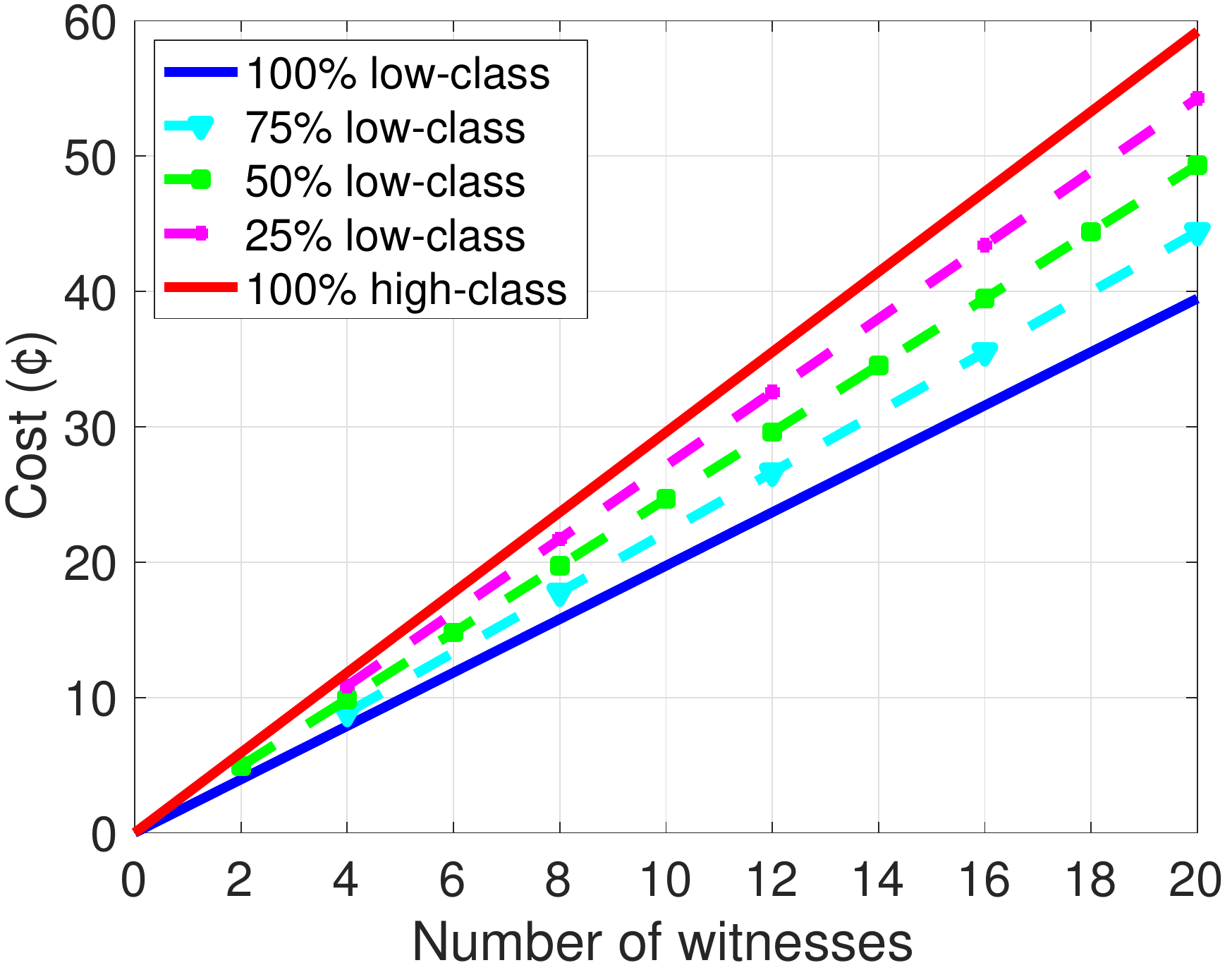}}\quad
				\label{fig:costtradeLowHigh}
			}
		}
		\vspace{-4mm}
		\caption{Dynamics of (a) verification error, and (b) total cost, as functions of the number of selected witnesses and the quality of their statements.}
		\vspace{-7mm}
		\label{fig:trend}
	\end{center}
\end{figure*}

%
%
%
%
%

In order to better analyze this optimization problem, we consider two classes of the witnesses including: (a) ``high-class'' witnesses which are powerful devices, affording more memory and power for witnessing service (giving quality statements at higher price), and (b) ``low-class'' witnesses that are relatively low power (giving less-accurate statements at lower price). 
Let us assume there exist $H$ high-class and $L$ low-class potential witnesses, offering the pair of error and cost as ($f_h$, $c_h$) and ($f_l$, $c_l$), respectively.
Now, our optimization problem becomes selecting the optimum number of witnesses from the two classes while keeping the total cost lower than a constant:
\begin{equation}
	\begin{aligned}
		\min \quad & {f_{h}^{H}f_{l}^{L}}\\
		\textrm{s.t.} \quad & c_{h}H + c_{l}L \leq C\\
		\quad & H ,L \in Z^{+} \label{opti}
	\end{aligned}
\end{equation}

The optimization is performed over variables $H$ and $L$, while false-positive probabilities ($f_i$'s) and associated costs ($c_i$'s) are known constants. We note that the objective function is a monotonically decreasing non-linear function of $H$ and $L$. 
Intuitively, the verification error (our objective) decreases by the number of witnesses selected, resulting in higher costs. This trend is magnified by the quality of combined witnesses. 

Fig.~\ref{fig:trend} shows the dynamics of verification error and total cost as a function of selected witnesses count  -- lines indicate the quality of selection (composition of high-/low-class witnesses). These values are computed by considering the following assumptions: $M=256$ (size of bloom filter); $N=150$ (total number of packets to be witnessed); $f_{h} = 0.15$ (false-positive rate of high-class witnesses); and $f_{l} = 0.35$ (false-positive rate of low-class witnesses). Given $M$, $f_{h}$ and $f_{l}$, we compute $n_{h} = 64$ and $n_{l} = 117$ from Eq.~\ref{bloom1}, resulting the number of statements per witness $m_h=3$ and $m_l=2$. 
We note that Ether (ETH) is the fuel for an Ethereum network. In order to interact with the Ethereum blockchain, user nodes need to pay to miner nodes for the computation of that transaction. That payment is calculated in gas  \cite{Gas} , and gas is always paid in ETH.
From our experimental setup (explained in \S\ref{sec:EthExp}), we found that submitting a 256-bit transaction requires spending $\alpha=2.77$\textcent~\cite{etherescan}. Therefore,  the price of offer for high-class and low-class witnesses, respectively, equals to $c_{h}=8.31$\textcent~and $c_{l}=5.54$\textcent~ (from Eq.~\ref{costformula}).


Here in Fig.~\ref{fig:trend}, we consider five scenarios ranging from selecting purely low-class witnesses (shown by solid blue lines) to mixes of low-/high-class witnesses (shown by dashed lines) to purely high-class witnesses (shown by solid red lines). It can be seen that the verification probability (one minus error) and the total cost (for the HSP) monotonically increase by the number of witnesses. Also, improving the quality of witnesses from ``100\% low-class" to ``100\% high-class'' accelerates the change rate (slope of lines) for both the verification error and the total cost. For example, given 10 witnesses, the verification error is $5 \times 10^{-5}$ when all witnesses are chosen from the low-class type (solid blue line in Fig.~\ref{fig:veritradeLowHigh}), resulting in a total cost of $20$\textcent~(solid blue line in Fig.~\ref{fig:costtradeLowHigh}). Improving the quality of witnesses to high-class fraction being 50\% (dashed green lines) and 100\% (solid red lines) would reduce  the error by 2 and 4 orders of magnitude, respectively, while incurring 25\% and 50\% additional cost. 

\begin{figure}[!t]
	\centering
	\includegraphics[width=0.40\textwidth,height=0.27\textwidth]{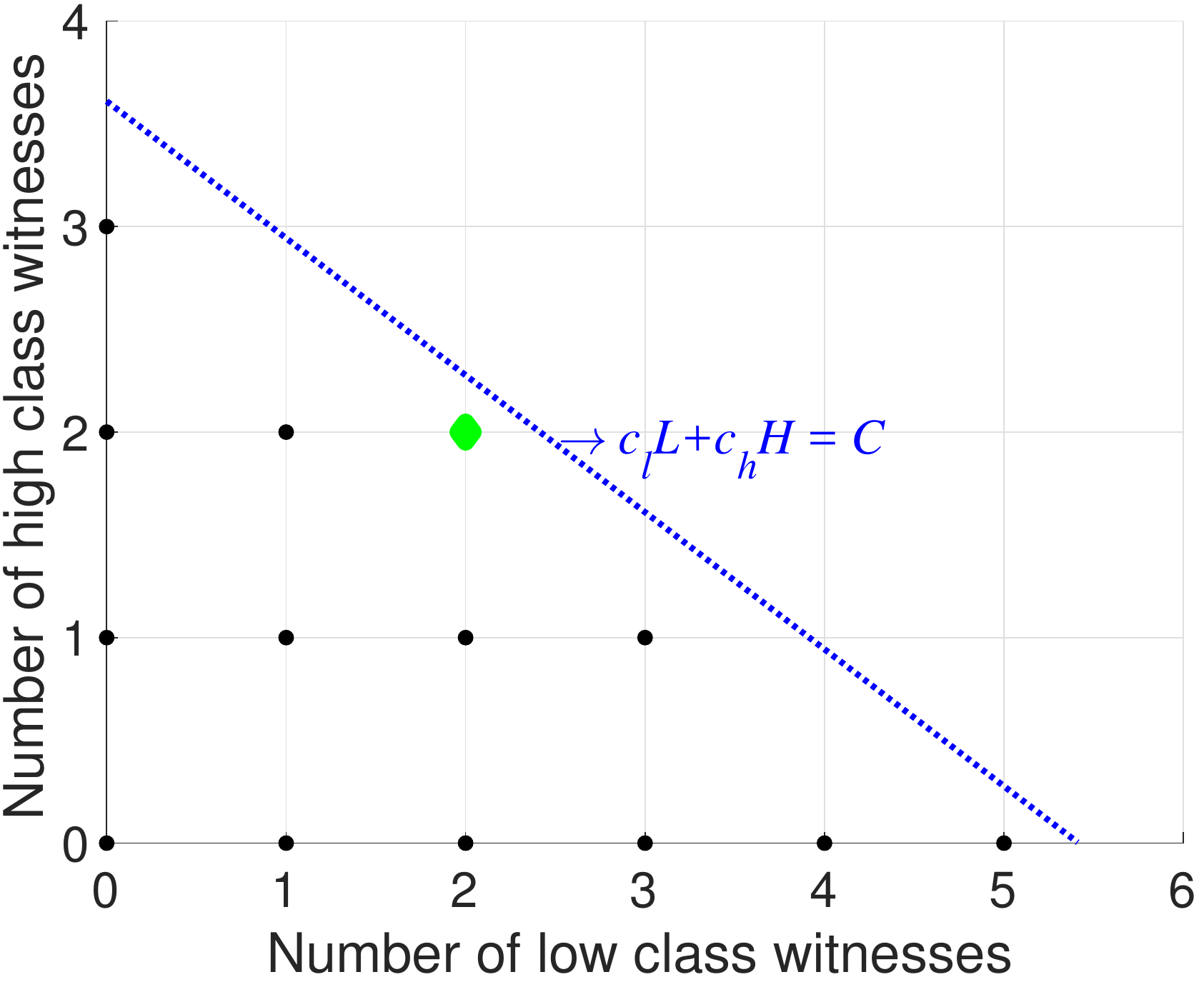}
	\vspace{-2mm}
	\caption{Example of feasible region and optimal point of the integer linear programming for $C=30$\textcent}
	\label{fig:MiLP}
\end{figure}


We note that our optimization problem (\ref{opti}) is an integer non-linear  programming. Since the objective function is a log-convex function and monotonically decreasing, its logarithmic transformation is also convex and monotonically decreasing \cite{luenberger1984linear}. This means that we can equally minimize the logarithmic transformation of the objective function in Eq.~\ref{opti}. Applying logarithmic transformation, multiplying the objective function by $\frac{-N\alpha}{M (ln(2))^{2}}$ and considering Eq.~\ref{costformula}, our optimization problem is expressed as a standard integer linear programming (ILP): {\color{blue}} 

\begin{equation}
	\begin{aligned}
		\max \quad & c_{h}H + c_{l}L\\
		\textrm{s.t.} \quad & c_{h}H + c_{l}L \leq C\\
		\quad & H ,L \in Z^{+} \label{ilp}
	\end{aligned}
\end{equation}



Restricting the variables to be positive ($H ,L \in Z^{+}$) in conjunction with having only one linear constraint with a negative slope will result in a triangle feasible region. Now, problem (\ref{ilp}) is a standard ILP, which does not have a closed-form solution \cite{bradley1977applied}. 
We illustrate in Fig.~\ref{fig:MiLP} the feasible region with a cost constraint $C=30$. The dotted blue line is our objective function and our optimal solution is the closest point (of the grid) to this line from the triangle region below it. In other words, the optimal solution is the combination of witnesses that yield a cost value close to this upper bound line. As highlighted by the green dot in Fig.~\ref{fig:MiLP}, our optimal solution is obtained by selecting two high-class witnesses ($H=2$) and two low-class witnesses ($L=2$).

Fig.~\ref{fig:optcost} shows the output of the optimization problem (Eq.~\ref{opti}) when the budget constraint varies from 0 to 120 cents.
Obviously, for a budget constraint less than $c_{l}=5.54$\textcent, it becomes infeasible to find the optimal solution (no witness can be selected). At a microscopic level, we observe that adding to the budget may at least increase the witnesses count or their quality.  At a macroscopic level, instead, relaxing the budget constraint would result in a larger number of witnesses (height of stacked bars) along with an improvement in their quality (height of red bars).

\begin{figure}[!t]
	\centering
	\includegraphics[width=0.45\textwidth,height=0.30\textwidth]{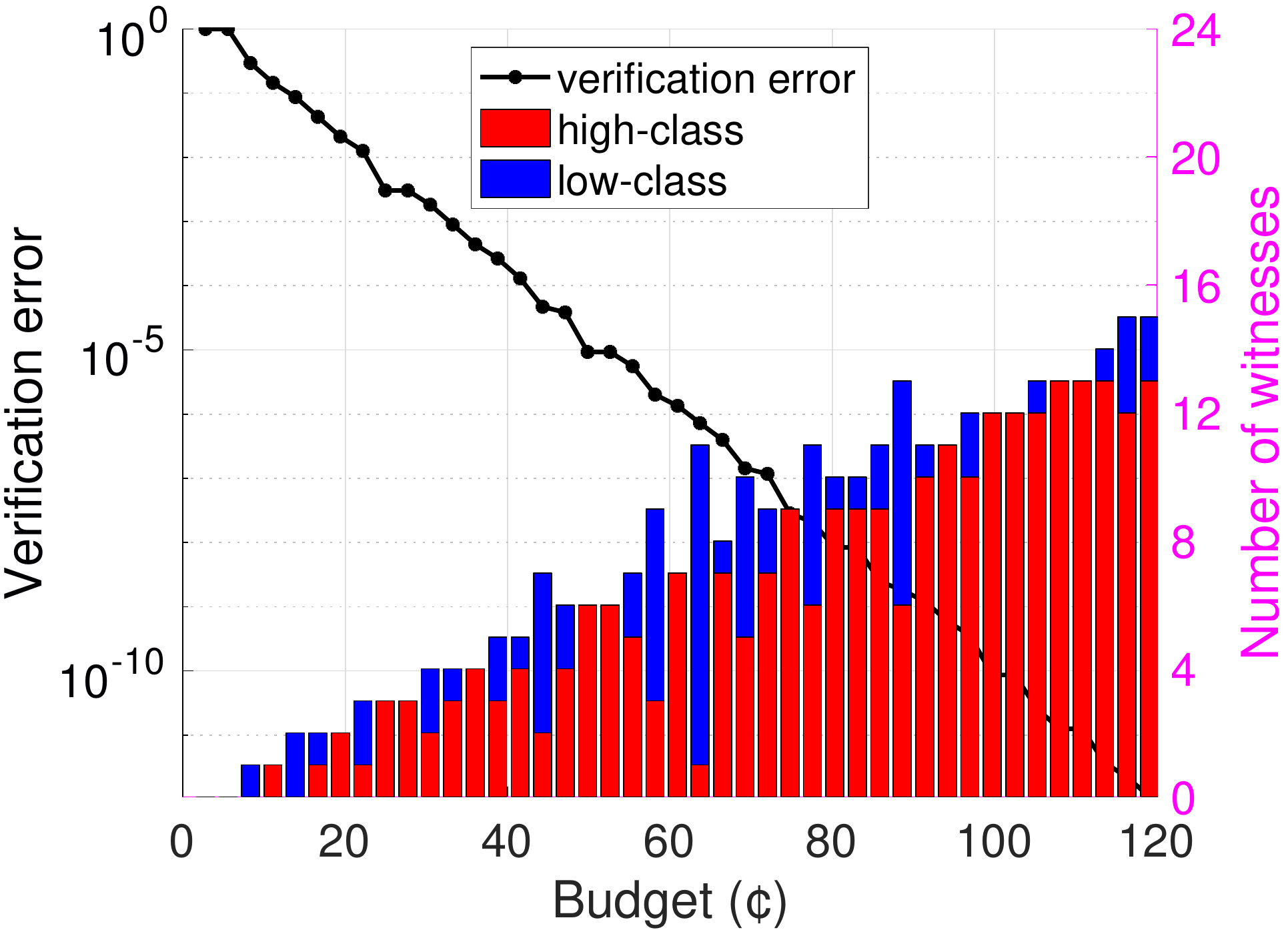}
	\vspace{-2mm}
	\caption{Optimal verification error and composition of witnesses versus budget constraint.}
	\label{fig:optcost}
\end{figure}

\section{Evaluation Results}\label{sec:eval}

We now evaluate the efficacy of our solution by applying it to real trace data.
We first set up a private Ethereum environment to deploy our smart contract and obtain the price of submitting transactions on the blockchain network. We next simulate our witness selection algorithm on trace data. Though our algorithm is designed for sensor networks at home or hospital premises, obtaining WiFi data from sufficient households to test the algorithm at scale is very challenging. To validate our algorithm at larger scale, we use traces taken from the WiFi network of a multi-story building on our university campus.


\subsection{Experiments with Ethereum Platform}\label{sec:EthExp}

We set up a private blockchain network on a machine (with 2.5 GHz Quad-Core Intel Core i7 and 6GB of memory) using the private Ethereum network \cite{GoEth} that is available for research purposes and private business use-cases. The private Ethereum network has the same APIs as the public Ethereum network \cite{privatenet}. To interact with the blockchain network, we use the Go-version Ethereum client (Geth v1.9.7) \cite{GoEth}. Also, we develop our smart contract as an app with four functions (\S\ref{sec:flow}) in Solidity v0.5.15. Lastly, we deploy the smart contract on the Ethereum blockchain using the Ethereum JavaScript API called {\myverb{web3.js}} v1.2.6 \cite{web3}. 
\begin{figure}[!t]
	\centering
	\includegraphics[width=0.45\textwidth,height=0.3\textwidth]{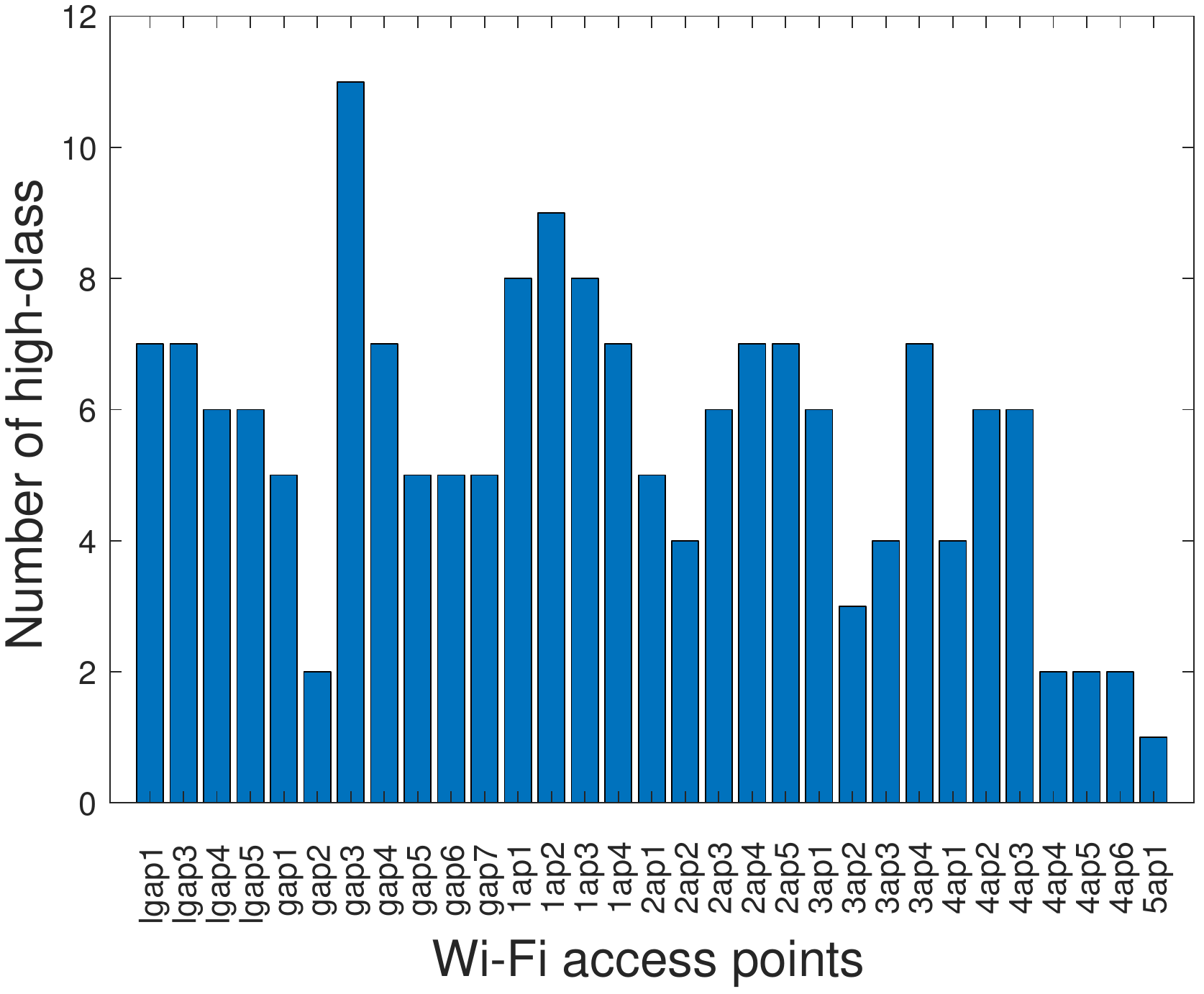}
	\vspace{-2mm}
	\caption{Number of high-class witnesses available around each of 31 WiFi access points.}
	\label{fig:fighigh}
\end{figure}

From our Ethereum testbed, we found that it requires $23,000$ gas in order to {\myverb{submit}} a witness statement of size 256 bits to the  blockchain. The dollar value of each gas fluctuates based on the dynamics of the crypto-currency market. 
At the time of our experiments (25 January 2020), 1,000,000 gas = 0.0075 Ethereum (ETH) and 1 ETH = 160.36\$ -- this means that 23000 gas is equal to $\alpha=2.77$\textcent~\cite{etherescan}.      
Note that witnesses will pay an amount (proportional to the size of transactions) for all transactions including {\myverb{offer}} and {\myverb{submit}}, but we only consider the cost of {\myverb{submit}} transactions in our witness selection algorithm.

\subsection{Evaluation in a Multi-Story Building}
To evaluate the efficacy of our scheme, we obtained WiFi trace data from our University IT department. The data contains session logs during a day (12:00am-11:59pm) for 31 WiFi access points (AP) located in a 5-story building. 
Each record contains device MAC address (note that we have hashed this to preserve anonymity); a unique AP name that clearly indicates the building name, floor level, and access
point ID; time at which the device associated to/disassociated from the AP (note that this is in minutes and therefore we do not have sub-minute accuracy); and avg throughput indicating
data rate during the session.

\begin{figure*}[t!]
	\begin{center}
		\mbox{
			\subfigure[Duration of WiFi sessions.]{
				{\includegraphics[width=0.3\textwidth,height=0.22\textwidth]{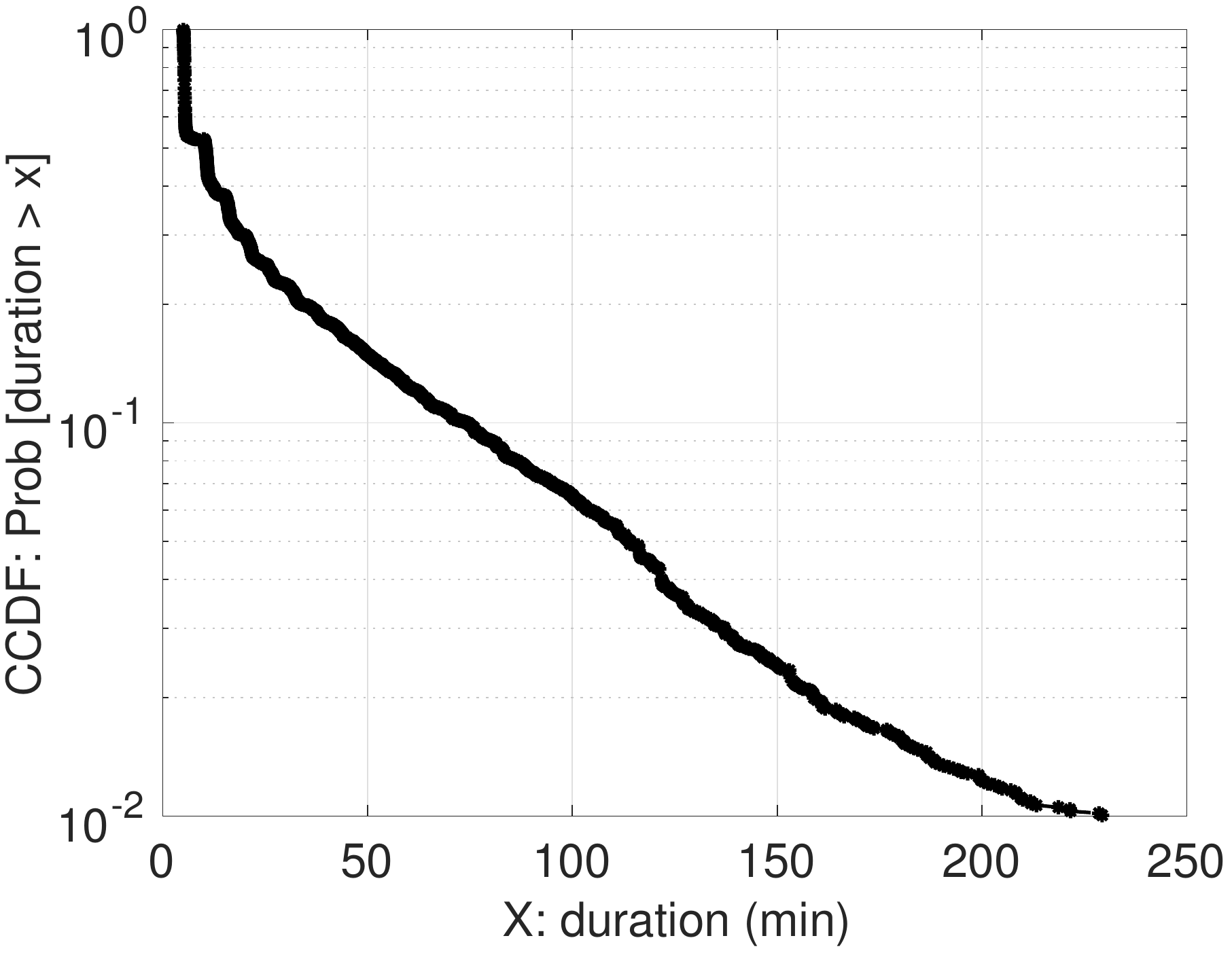}}\quad
				\label{fig:CCDFduration}
			}
		}
		\hspace{-5mm}
		\mbox{
			\subfigure[Cost of selecting \underline{all} available witnesses.]{
				{\includegraphics[width=0.3\textwidth,height=0.22\textwidth]{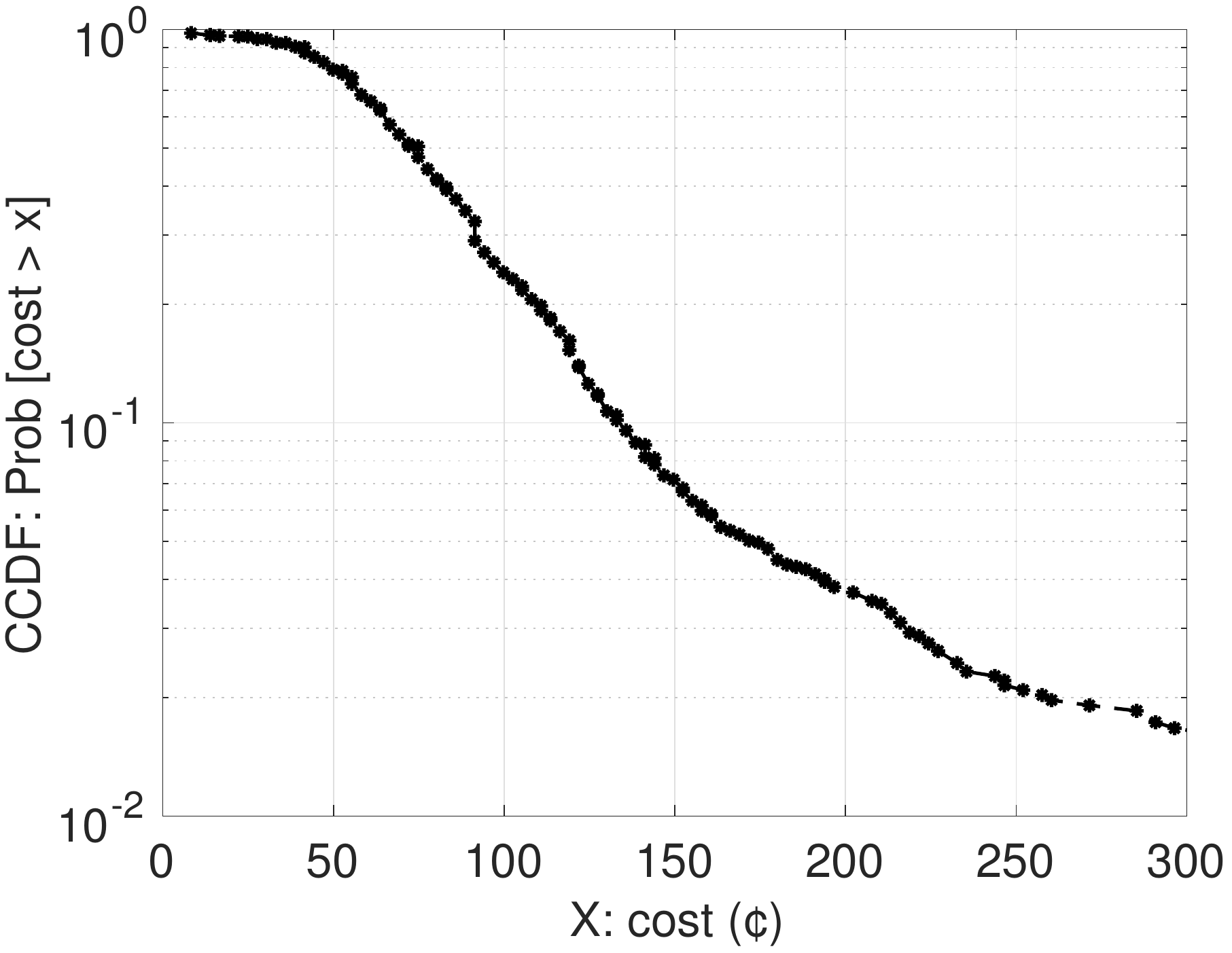}}\quad
				\label{fig:CCDFprice}
			}
		}
		\hspace{-5mm}
		\mbox{
			\subfigure[Maximum epoch cost per zone.]{
				{\includegraphics[width=0.3\textwidth,height=0.22\textwidth]{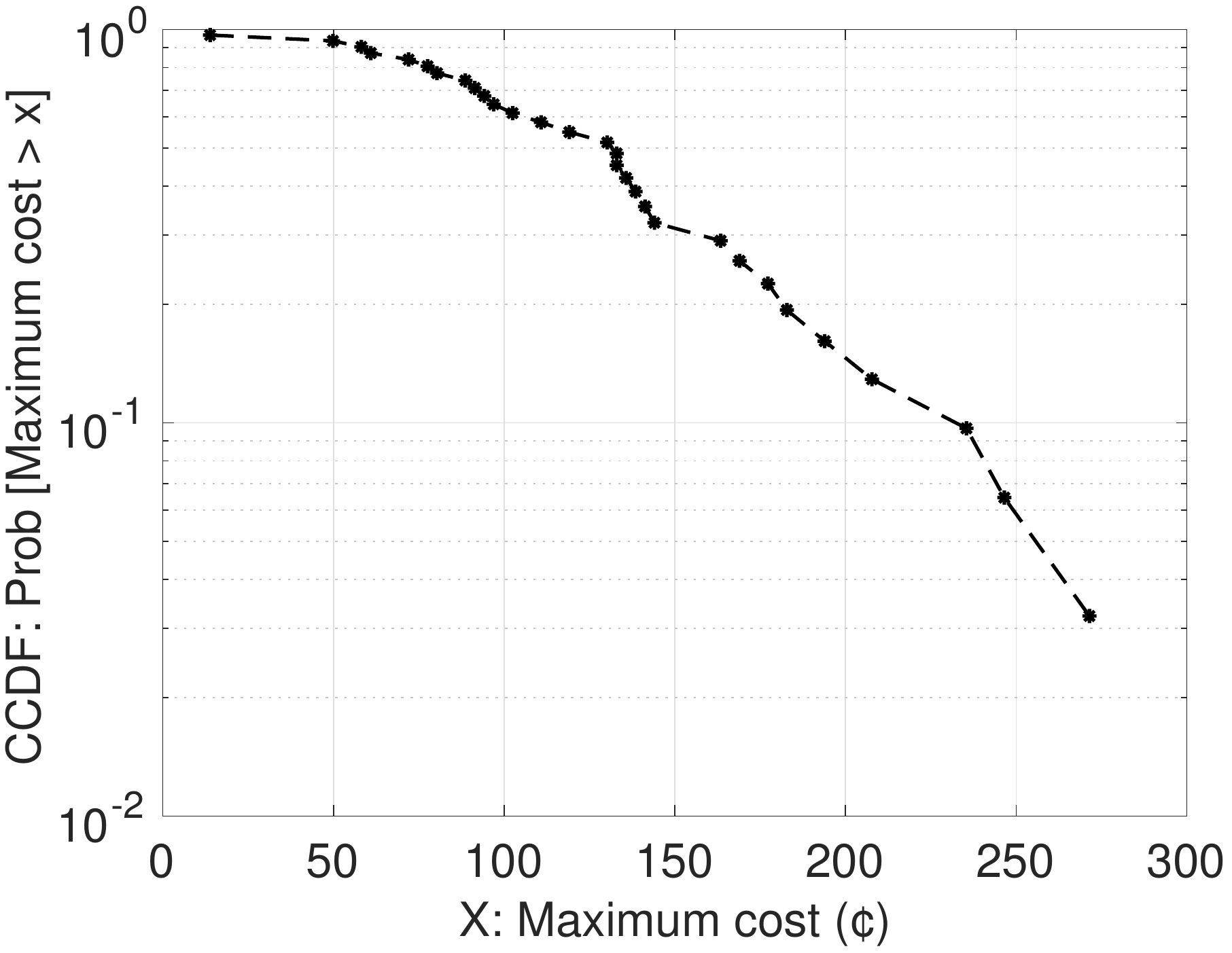}}\quad
				\label{fig:CCDFpriceMAX}
			}
		}
		
		\vspace{-2mm}
		\caption{CCDF of: (a) duration of WiFi sessions in our trace data, (b) cost of selecting ``all'' available witnesses across all epochs and WiFi zones, and (c) maximum of epoch cost per zone.}
		\vspace{-2mm}
		\label{fig:CCDF}
	\end{center}
\end{figure*}

\begin{figure*}[t!]
	\begin{center}
		\mbox{
			\subfigure[Available witnesses.]{
				{\includegraphics[width=0.3\textwidth,height=0.22\textwidth]{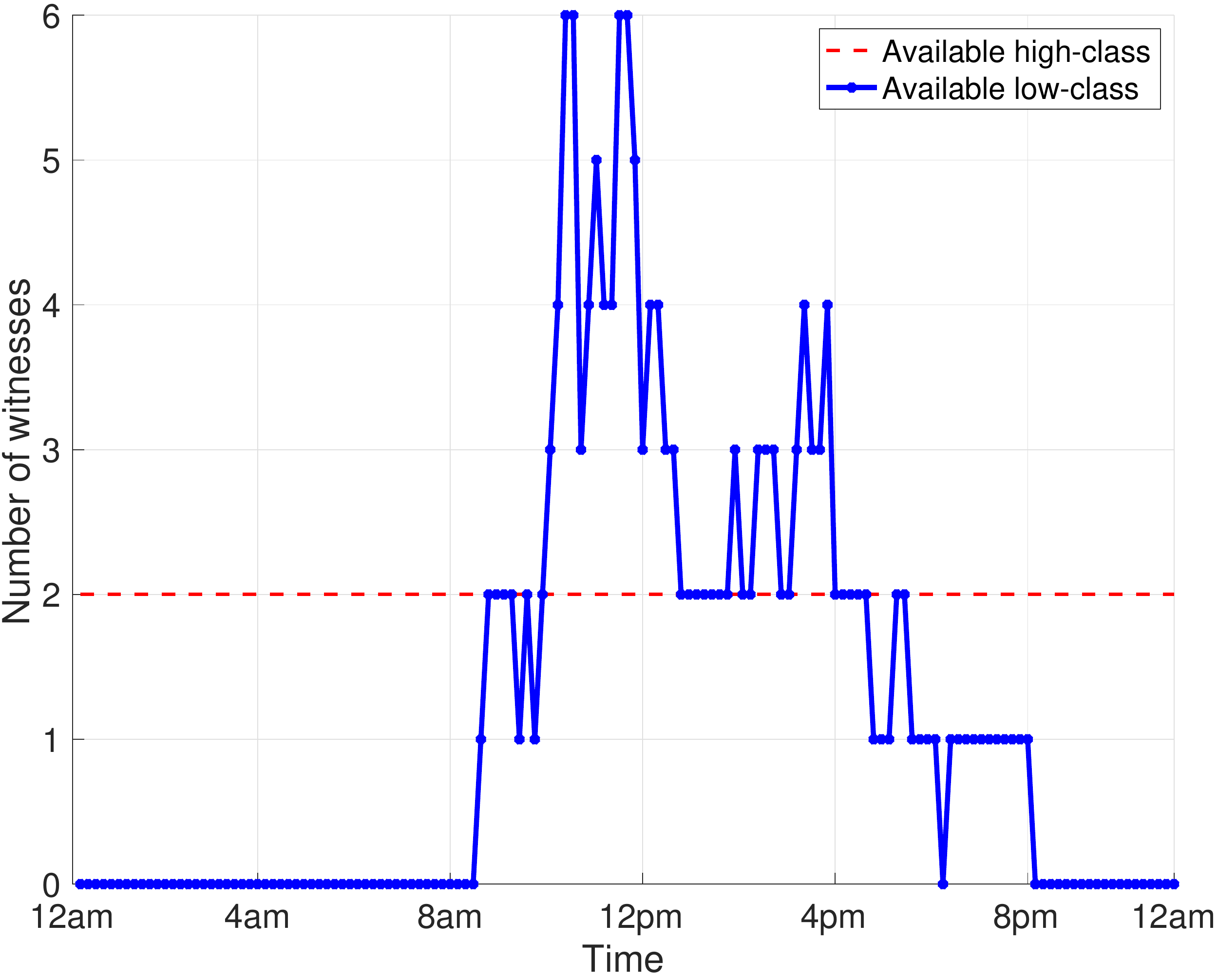}}\quad
				\label{fig:noOPTwitness}
			}
		}
		\hspace{-5mm}
		\mbox{
			\subfigure[Cost of witnessing.]{
				{\includegraphics[width=0.3\textwidth,height=0.22\textwidth]{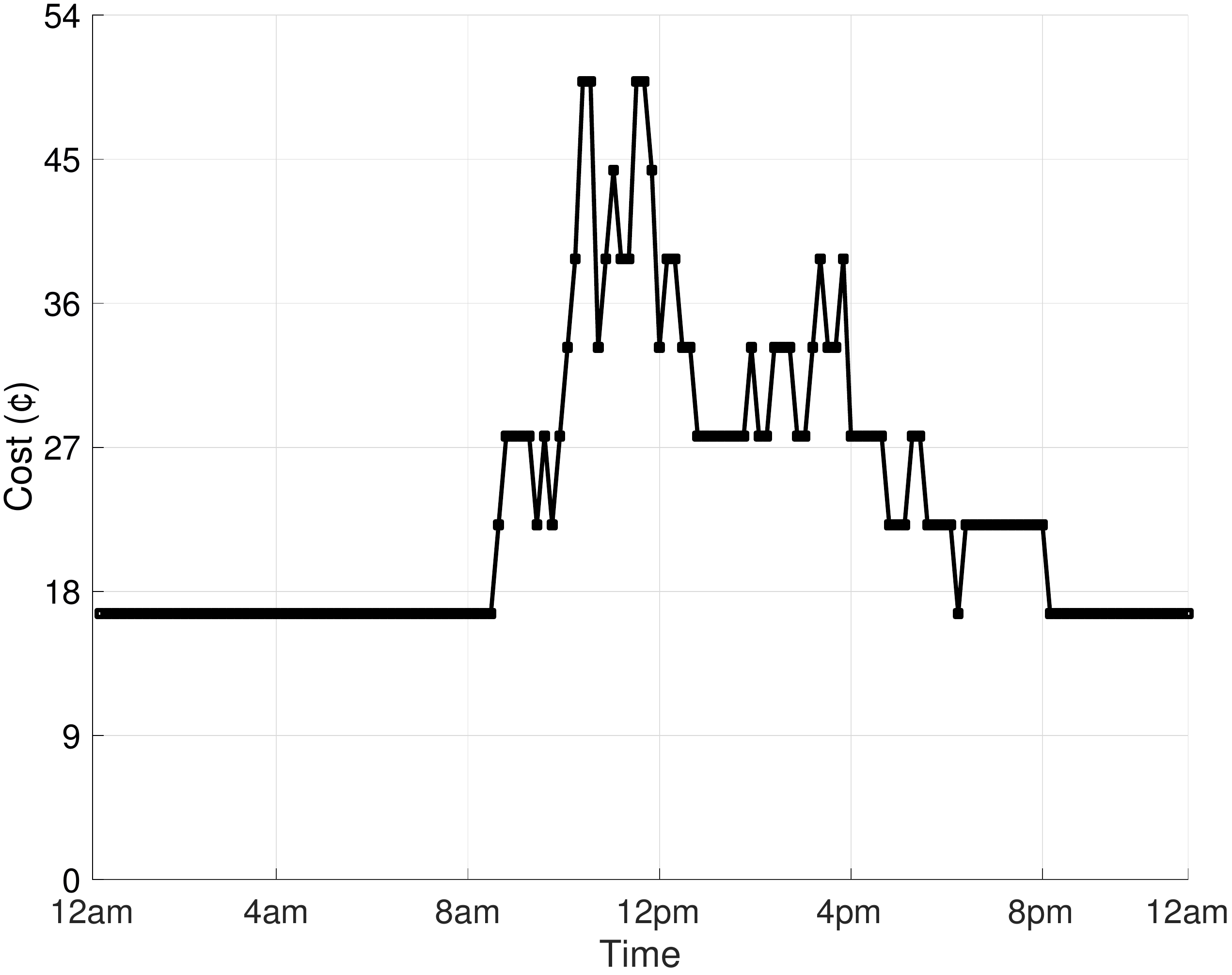}}\quad
				\label{fig:noOPTcost}
			}
		}
		\hspace{-5mm}
		\mbox{
			\subfigure[Verification error of witnessing.]{
				{\includegraphics[width=0.3\textwidth,height=0.22\textwidth]{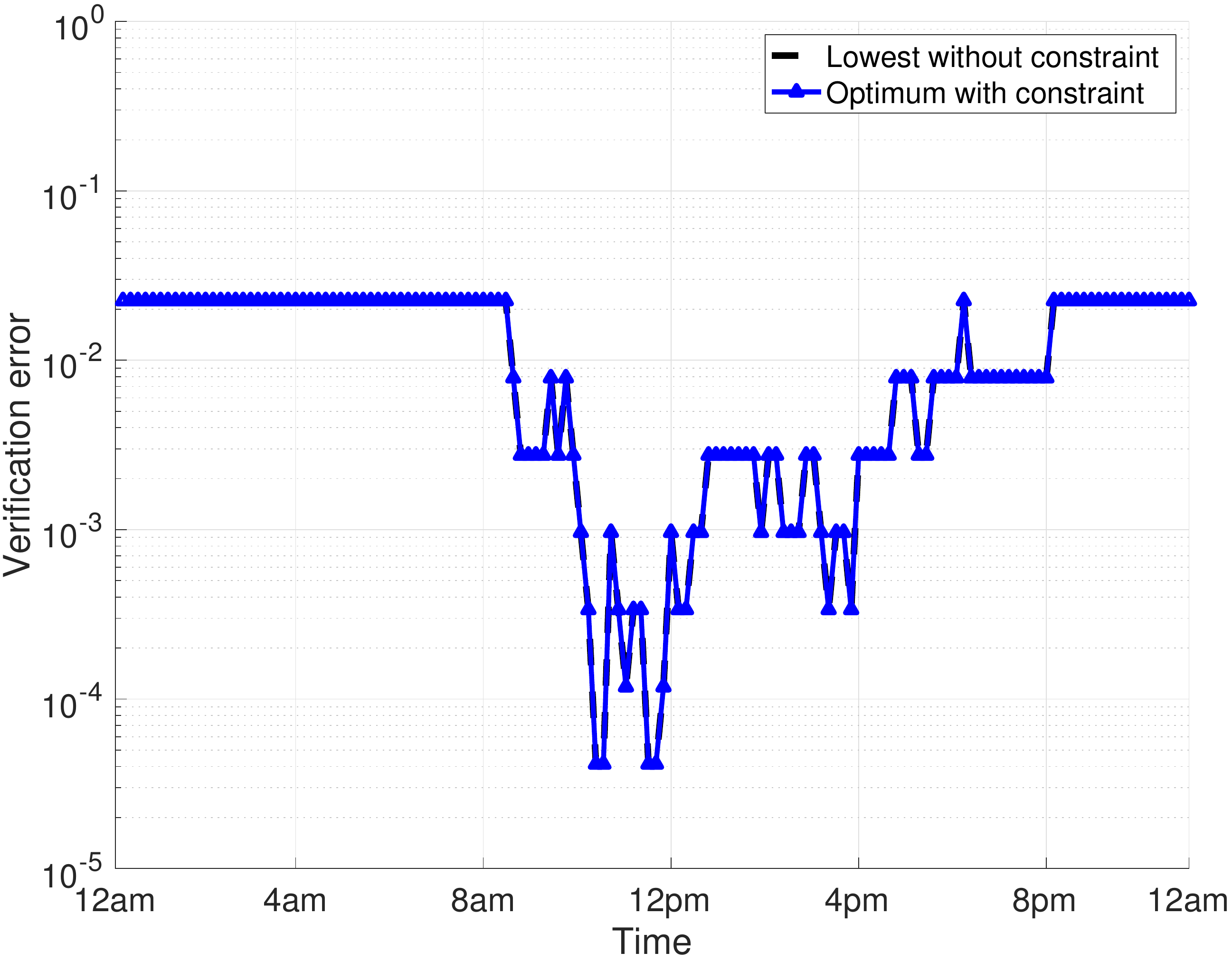}}\quad
				\label{fig:noOPTerr}
			}
		}
		\vspace{-2mm}
		\caption{Dynamics of witnessing for a low-density zone {\textit{4ap5}} when optimization is \underline{not} needed: (a) count of available witnesses, (b)) cost of witnessing, and (c) verification error of witnessing.}
		\vspace{-2mm}
		\label{fig:noOPT}
	\end{center}
\end{figure*}

To simulate our scheme, we assume that each AP (static and powerful) represents a high-class witness and a user device (mobile with limited compute power) represents a low-class witness. Since APs are spread across the building, we call the coverage area of each AP a ``zone'' from which user devices connect to the AP. Obviously wireless zones overlap, and hence for a given AP there exist a number of ``neighbor APs'' within its close proximity.
A witnessing scenario in this environment is as follow. We consider our target sensor in an AP zone with several high-class witnesses (neighbor APs) and a number of low-class witnesses (user devices connected to the AP). Fig.~\ref{fig:fighigh} shows the distribution of high-class witnesses across 31 zones -- each zone corresponds to a WiFi AP in the building. It can be seen that each zone comprises 5 high-class witnesses on average. This number varies in certain zones -- for example zone \textit{gap3} located at ground level covers 11 APs, or zone \textit{5ap1} located at level5 accommodates only one AP.

Obviously, high-class witnesses (neighbor APs) consistently stay present within their corresponding zones, but low-class witnesses (user devices) are mobile and hence their availability changes over time. To get a sense of the availability of low-class witnesses, we plot in Fig.~\ref{fig:CCDFduration} the complementary cumulative distribution function (CCDF) of all sessions duration in our trace. Of a total of 8263 WiFi sessions, more than half ($52$\%) had a duration more than 10 minutes. Therefore, we consider a witnessing epoch to last 10 minutes. A target wearable sensor is assumed \cite{siddiqi2019secure} to transmit (on average) 15 packets per minute, and hence a total of $N=150$ packets will be witnessed during a witnessing epoch. 

In our simulation, we divide a day into 144 distinct epochs of 10-minute. For each epoch, we only consider those user devices as potential low-class witnesses that remain connected to the same zone for the whole duration of that epoch -- devices which switch their zone during the epoch are filtered out. This means that potential witnesses are required to be in the vicinity of the target sensor (present in the zone) during an epoch.   
With this condition, $27$\% of sessions on average are removed per epoch-zone -- note that in three-quarters of epoch-zones less than $40$\% of sessions are filtered out, while for a third of epoch-zones all sessions persist during the epoch (no session was filtered out).

\begin{figure*}[t!]
	\begin{center}
		\mbox{
			\subfigure[Available witnesses.]{
				{\includegraphics[width=0.45\textwidth,height=0.27\textwidth]{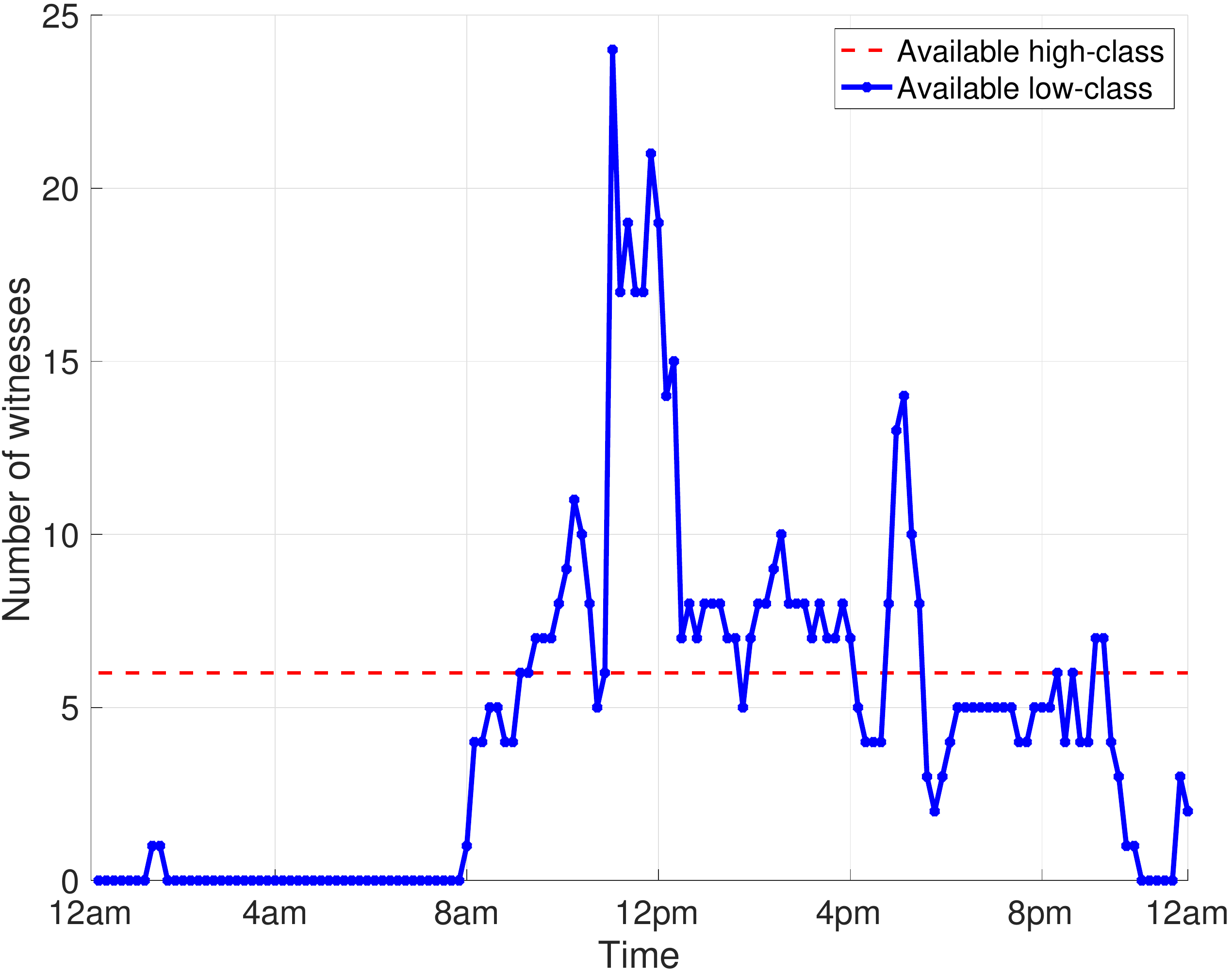}}\quad
				\label{fig:OPTmax}
			}
		}
		\hspace{-3mm}
		\mbox{
			\subfigure[Selected witnesses.]{
				{\includegraphics[width=0.45\textwidth,height=0.27\textwidth]{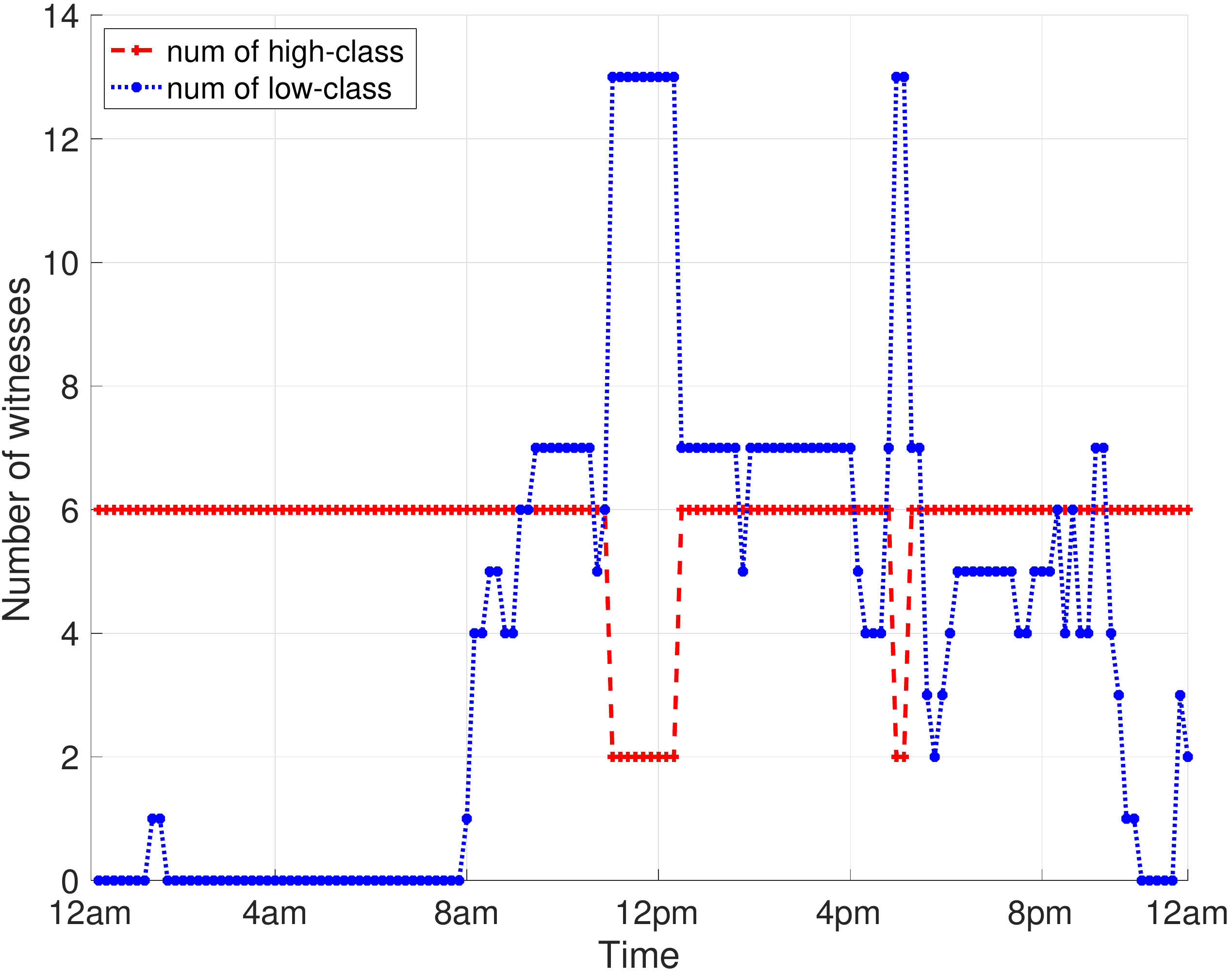}}\quad
				\label{fig:OPTwitness}
			}
		}
		\hspace{-3mm}
		
		\mbox{
			\subfigure[Cost of witnessing.]{
				{\includegraphics[width=0.45\textwidth,height=0.27\textwidth]{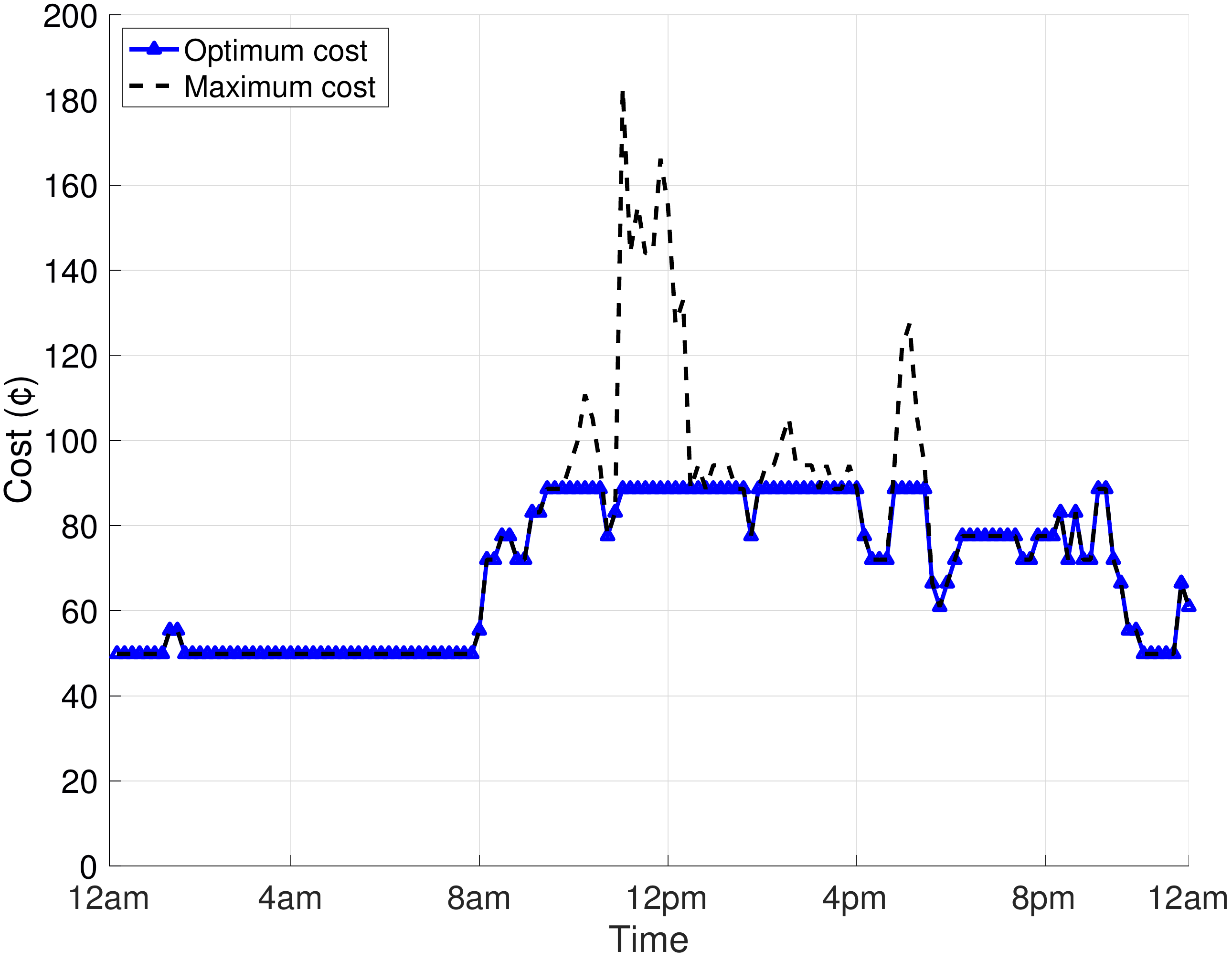}}\quad
				\label{fig:OPTcost}
			}
		}
		\hspace{-3mm}
		\mbox{
			\subfigure[Verification error of witnessing.]{
				{\includegraphics[width=0.45\textwidth,height=0.27\textwidth]{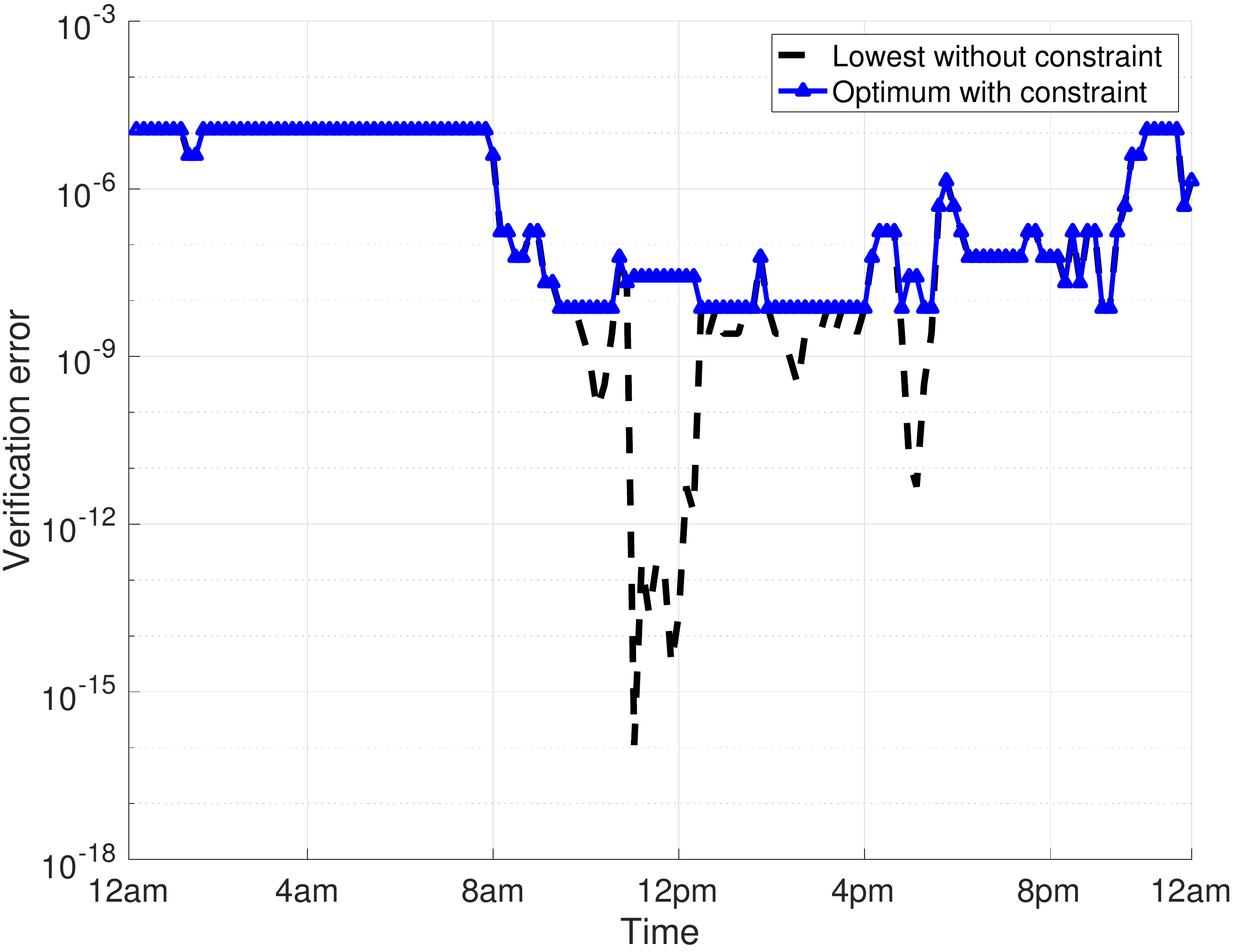}}\quad
				\label{fig:OPTerr}
			}
		}
		\vspace{-2mm}
		\caption{Dynamics of witnessing for a high-density zone {\textit{4ap2}} when optimization is needed: (a) count of available witnesses, (b) count of selected witnesses, (c) cost of witnessing, and (d) verification error of witnessing.}
		\vspace{-5mm}
		\label{fig:OPT}
	\end{center}
\end{figure*}

Moreover, we note that cost constraints and pricing strategy chosen by the HSP can be influenced by a number factors such as the importance of data transmitted by the target sensor, the matter of urgency for detecting security breaches, the availability of potential witnesses, or the number of sensors to be witnessed at scale. 
It is important to note that having a generous budget for the HSP may result in selection of all available witnesses, making the optimization unnecessary. On the other hand, a very tight budget can make it infeasible for the optimization algorithm to select even one low-class witness. For this paper, we consider a simple pricing strategy for the HSP whereby a fixed budget is pre-decided  for every epoch, and all epochs are treated equally -- more sophisticated strategies can explored in future works. 

Recall from \S\ref{sec:verification}, selecting a witness would cost $c_{h} = 8.31$\textcent~for the high-class and $c_{l} = 5.54$\textcent~for the low-class, given $\alpha=2.77$\textcent~(computed from our Ethereum testbed in \S\ref{sec:EthExp}) and $N=150$ packets transmitted by the target sensor over a 10-minute epoch. 
Note that the number of witnesses (both high-class and low-class) during each epoch is known, and hence the incurred cost incurred of selecting all available witnesses (in a given zone) can be computed. Fig.~\ref{fig:CCDFprice} shows the CCDF of cost per epoch during day-time (\ie 8am-5pm when user devices are likely present) across all zones. We observe that for a given zone/epoch there is an $80$\% chance to have the cost more than 50\textcent, when the HSP chooses to select all available witnesses in the sensor environment. Note that some zones get more crowded than others depending on their location (meeting rooms, study spaces, offices, research labs, or entry/exit points) in the building. We computed the maximum cost of each zone across epochs of the day. Fig.~\ref{fig:CCDFpriceMAX} shows the CCDF of maximum cost per zone. We observe that for a third of zones (low-density ones) the cost is always below 90\textcent, hence for these zones the HSP may prefer to go with all available witnesses in the environment without employing the optimization algorithm. We, therefore, set the budget constraint to a fixed value equals to 90\textcent~in our simulations, highlighting the fact that the optimization may be needed in two-third of the zones (high-density ones) where there likely be a large number of mobile devices available to perform witnessing.

Given the budget constraint, we now simulate our scheme during the whole 24 hours in two representative zones: a low-density (AP \textit{4ap5}) and a high-density (AP \textit{4ap2}). Fig.~\ref{fig:noOPT} shows the dynamics of witnessing (number of witnesses, cost, and verification error) for the low-density zone. It is seen in Fig.~\ref{fig:noOPTwitness} that this zone accommodates two high-class witnesses (shown by a flat dashed red line) and up to 6 low-class witnesses (mostly 1-3 during working hours as shown by dotted blue lines). For this low-density environment, no optimization is needed since the total cost of selecting all available witnesses is well below the budget constraint 90\textcent, and hence all available witnesses are selected. We observe in Fig.~\ref{fig:noOPTcost} that the highest cost is about 54\textcent~ for epochs between 10am-11:30am when the total count of witnesses reaches to its maximum of 8. As a result, the lowest verification error becomes identical to the optimal verification, as shown in Fig.~\ref{fig:noOPTerr} -- the lowest error of  $10^{-4}$ can be achieved at the cost of 54\textcent~per epoch during the peak time of this zone.


Moving to the high-density zone which hosts 6 high-class witnesses and 5-24 low-class witnesses during day time, as shown in Fig.~\ref{fig:OPTmax}. The HSP now needs to run the optimization algorithm for selecting the optimal combination of witnesses available in the environment. We observe that during early morning (12am-9am), selecting all available witnesses results a cost less than the constraint 90\textcent, as shown in Fig.~\ref{fig:OPTcost}. Following 9am, it is seen that the cost is saturated at 90\textcent~ as a result of optimization. 
Focusing on optimization results, we observe that two-third of high-class witnesses are left out due to abundance of low-class witnesses during busy epochs (around 12pm and closer to 5pm), as shown in Fig.~\ref{fig:OPTwitness}. In this scenario, the optimum verification error would be higher than the lowest error as shown in Fig.~\ref{fig:OPTerr} -- during the peak time the best error $10^{-8}$ is achieved by selecting 2 high-class and 13 low-class witnesses.


\section{Conclusion}\label{sec:conc}
In this paper, we have proposed an on-demand and distributed scheme to verify healthcare IoT data. Our architecture
provides the motivation and means to engage via smart contracts on a blockchain:  Health authorities can request witness statements needed for data verification of target sensors; local witness devices can monetize their statements without compromising their privacy.
We developed an optimization algorithm for health authorities to select an optimal collection of available witnesses to achieve the best verification probability subject a budget constraint.
We simulated our algorithm on real data captured from Wi-Fi connections in a multi-story campus building to show that a verification probability of more than 99.99\% can be achieved at cost of less than two dollars for one-hour witnessing service. This work is the first step towards on-demand witnessing of sensors network data, applicable to real-world scenarios.

 \ifCLASSOPTIONcaptionsoff
   \newpage
 \fi

\bibliographystyle{IEEEtran}
\balance
\bibliography{witnessBC}

\end{document}